\documentclass[12pt,a4paper]{article}
\usepackage{xcolor,paralist}
\usepackage[pdftex,colorlinks=true,hypertexnames=false]{hyperref}
\definecolor{darkblue}{rgb}{0,0,.6}
\hypersetup{citecolor=darkblue,linkcolor=darkblue,urlcolor=darkblue}
\usepackage{amssymb,enumitem,booktabs,subfig,setspace,bm,longtable,dsfont,orcidlink}
\usepackage[final,nopatch=footnote]{microtype}
\usepackage[top=1in, bottom=1in, left=1in, right=1in]{geometry}
\usepackage{natbib}
\usepackage[thinc]{esdiff}
\usepackage{comment,mathtools}
\usepackage{dsfont}
\usepackage{url}
\usepackage{amsmath}
\usepackage{amsfonts}
\usepackage{epsfig}
\usepackage{graphics,algorithm,algpseudocode}
\usepackage[normalem]{ulem}
\usepackage{palatino,eulervm}
\usepackage{tikz}
\usetikzlibrary{arrows.meta,decorations.pathreplacing}
\usepackage{graphicx}
\usepackage{lscape}
\usepackage{rotating}
\usepackage[linewidth=1pt]{mdframed}
\allowdisplaybreaks[4]
\DeclareMathOperator*{\argmin}{arg\,min}

\providecommand{\U}[1]{\protect\rule{.1in}{.1in}}

\definecolor{colCPD}{HTML}{FF0000}
\definecolor{colMIS}{HTML}{0000FF}
\graphicspath{{plots/}}
\newsavebox\CBox

\usepackage{amsthm,thmtools}
\usepackage{mathrsfs}

\makeatletter
\def\th@newremark{\th@remark\thm@headfont{\bfseries}}
\makeatletter
\theoremstyle{newremark}

\declaretheoremstyle[spaceabove=6pt, spacebelow=6pt, headfont=\bfseries, notefont=\mdseries, notebraces={(}{)}, bodyfont=\normalfont, postheadspace=0.5em]{mystyle}

\newcommand{\X}{\mathcal{X}}
\newcommand{\Y}{\mathcal{Y}}
\newcommand{\Z}{\mathcal{Z}}
\bibpunct{(}{)}{;}{a}{,}{ and }

\makeatletter
\newcommand*{\addFileDependency}[1]{
\typeout{(#1)}
\@addtofilelist{#1}
\IfFileExists{#1}{}{\typeout{No file #1.}}
}\makeatother

\newcommand{\Rlogo}{\protect\includegraphics[height=1.8ex,keepaspectratio]{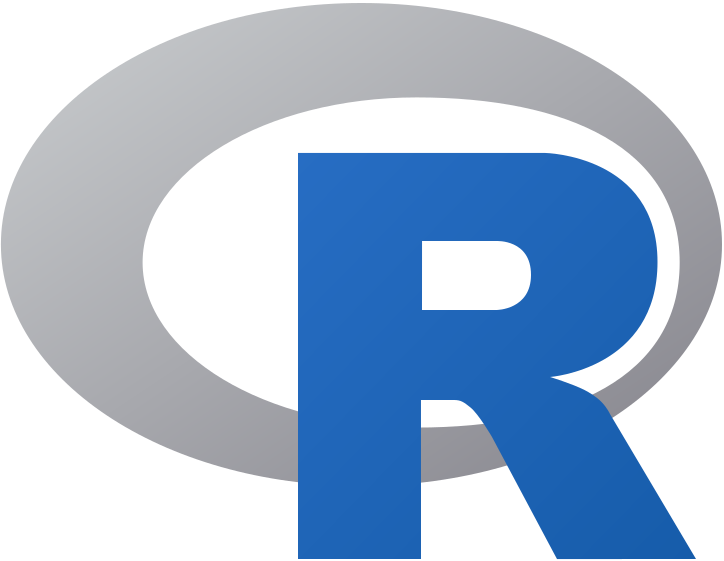}}

\begin{document}

\title{Conformal prediction for high-dimensional functional time series: Applications to subnational mortality}
\author{
\normalsize Han Lin Shang \orcidlink{0000-0003-1769-6430} \qquad Chang Liu \footnote{Corresponding address: Department of Actuarial Studies and Business Analytics, Level 7, 4 Eastern Road, Macquarie University, Sydney, NSW 2109, Australia. Telephone: +61(4) 3519 3674; Email: chang.liu45@students.mq.edu.au} \orcidlink{0009-0007-6909-3195}\\
\normalsize Department of Actuarial Studies and Business Analytics \\
\normalsize Macquarie University 
}

\date{}

\maketitle

\begin{abstract}
In statistics, forecast uncertainty is often quantified using a specified statistical model, though such approaches may be vulnerable to model misspecification, selection bias, and limited finite-sample validity. While bootstrap procedures may alleviate some of these concerns, their validity is not guaranteed and they are often computationally demanding. Instead, we take a model-agnostic and distribution-free approach, namely conformal prediction, to construct pointwise prediction intervals in high-dimensional functional time series. Among a rich family of conformal prediction methods, we study split, non-conformity score, sequential predictive, and adaptive conformal prediction. In split and non-conformity score conformal prediction, the data are divided into training, validation, and test sets, and the interval width is calibrated on the validation set so that the empirical coverage probability matches the nominal level, by tuning a coefficient of the residual standard deviation and by taking an empirical quantile of the non-conformity scores, respectively. Sequential predictive conformal prediction removes the need for a validation set by updating predictive quantiles sequentially via an autoregressive process, whereas adaptive conformal prediction updates the significance level over time according to the coverage of past intervals. Using subnational age- and sex-specific $\log$ mortality data from Japan and Canada, we compare the finite-sample forecast performance of the conformal prediction methods considered, using empirical coverage probability and the mean interval score.

\vspace{.1in}
\end{abstract}

\newpage
\setstretch{1.54}

\section{Introduction}\label{sec:1}

A fundamental problem in modeling and forecasting time series of random functions or objects is quantifying forecast uncertainty. Such uncertainty is often manifested probabilistically through the construction of pointwise prediction intervals or uniform prediction bands. The prediction interval or band allows one to assess future uncertainty levels, plan different strategies for a range of possible outcomes indicated by the interval forecasts, compare forecasts from different methods more thoroughly, and explore different scenarios based on various assumptions \citep{Chatfield00}. In the functional time series literature, \cite{HU07} estimated the total variance and constructed prediction intervals based on a normality assumption. To avoid the parametric distributional assumption, \cite{PS23} proposed a sieve bootstrap method to construct prediction bands for a single functional time series. The sieve bootstrap generates bootstrap samples of stationary functional time series and their forecasts via a vector autoregression representation, and it can calibrate the model misspecification error, which does not vanish asymptotically.

While most of the existing literature focuses on a small (fixed) number of functional time series, it is increasingly common to collect large numbers of them across diverse fields. For instance, in climatology, temperature curves are routinely recorded at hundreds of weather stations; in finance, return curves are typically available for thousands of stocks. These applications are examples of high-dimensional functional time series (HDFTS), which have received increasing attention in recent years. An HDFTS can be denoted by $\Z_{t,s}(u), s=1,\dots, N, t=1,\dots, T$ with $\Z_{t,s} = (\Z_{t,s}(u), u\in \mathcal{C}_s)\in \mathcal{H}_s$, where $\mathcal{H}_s$ is a Hilbert space defined as a set of measurable and square-integrable functions on a compact set $\mathcal{C}_s$. A distinguishing feature of HDFTS is that the number of cross-sectional units may exceed the number of time periods. In our application, each of the Japanese female and male populations comprises $47$ series, one per prefecture, observed over 49 years, so that $N=47$ and $T=49$.

In the functional time series literature, the study of HDFTS is in its infancy. \cite{HNT23} presented a theoretical representation within a functional factor model framework, while \cite{LLS24} developed an information criterion for detecting and estimating change points and for clustering them into common change points. For estimating parameters in an HDFTS model,  \cite{TNH23} proposed a functional factor model with functional factor loadings and real-valued factors, whereas \cite{GQW+26} derived a different factor model with real-valued factor loadings and functional factors; by unifying the two factor models, \cite{LLS+26} developed a unified factor model with functional factor loadings and functional factors. Because of cross-sectional and temporal dependencies in the HDFTS, \cite{HG18} combined principal component scores derived from different populations and subsequently performed a separate real-valued principal component analysis on the aggregated scores. \cite{GSY19} adopted a two-stage procedure that integrates truncated functional principal component analysis with a scalar factor model applied to the resulting panel of scores. \cite{JSS24, JSS25} developed a two-way functional analysis of variance to decompose HDFTS into deterministic components, namely grand, row, and column effects, and time-varying residual processes. \cite{TSY22} implemented a multilevel functional time-series model to extract common and series-specific patterns, and developed techniques for clustering multiple functional time series.

To quantify forecast uncertainty, the main aim of this paper is to propose an approach for constructing prediction intervals for HDFTS using a conformal prediction approach. Because they are computationally efficient, we study the split conformal prediction of \cite{LSR+18}; the non-conformity score prediction, which sets the interval by the empirical quantile of width-standardized non-conformity scores \citep{RPC19, KTL23} and so requires no tuning; the adaptive conformal prediction of \cite{GC21}, which updates the significance level over time according to the coverage of past intervals \citep[see also][]{ZFG+22, GC24}; and the sequential predictive conformal prediction of \cite{XX23}, designed for time series under distribution shift. Among these conformal prediction methods, we recommend adaptive conformal prediction at short horizons, as it attains the most accurate coverage, and sequential predictive conformal prediction at longer horizons, as it does not require calibration on a validation set and attains the smallest mean interval score (MIS) with comparable coverage. To the best of our knowledge, this is the first attempt to apply conformal prediction in an HDFTS setting. This work can be seen as an extension of \cite{Shang26b} from a single functional time series to HDFTS, with the inclusion of the non-conformity scores and adaptive conformal prediction.

The remainder of the paper is organized as follows. In Section~\ref{sec:2}, we introduce a motivating example, namely Japanese subnational age- and sex-specific log mortality rates. Section~\ref{sec:3} presents a two-stage method: a one-way functional ANOVA (Section~\ref{sec:3.1}) followed by the functional factor model of \cite{LLS+26} applied to the residual processes (Section~\ref{sec:3.2}). Since the functional ANOVA decomposition can exactly recover an HDFTS, there is no loss of information. From this two-stage decomposition, we obtain forecasts of the real-valued factors and, in turn, of the HDFTS in Section~\ref{sec:3.3} and summarize all steps in an algorithm in Section~\ref{sec:3.4}. In Section~\ref{sec:4}, we present the conformal prediction methods, tailored for HDFTS. In Section~\ref{sec:5}, we describe the forecast evaluation setup and error metrics. The interval forecast accuracy is evaluated and compared between the conformal prediction methods in Section~\ref{sec:6}. Section~\ref{sec:7} concludes with some ideas on how the methodology can be further extended. In the Appendix, we present a sensitivity analysis by applying the conformal prediction methods to Canadian subnational age- and sex-specific mortality rates to further validate our results.

\section{Japanese subnational age- and sex-specific mortality rates}\label{sec:2}

\begin{figure}[!htb]
\centering
\includegraphics[width=5.1cm]{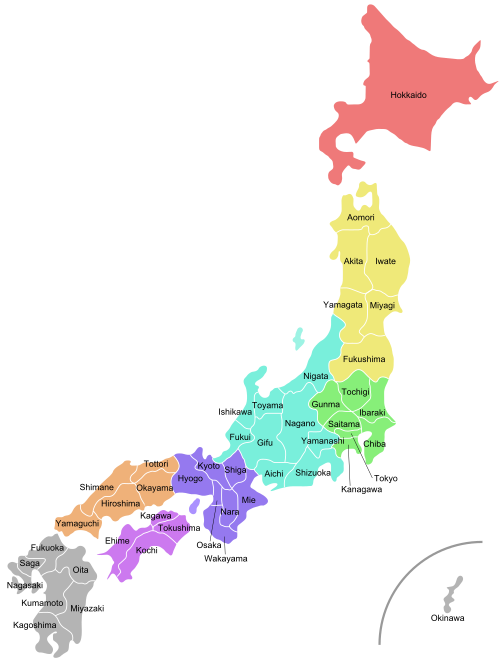}
\caption{\small A map of 47 Japanese prefectures ordered geographically from the northernmost region (Hokkaido) to the southernmost region (Okinawa).}\label{fig:1}
\end{figure}

We demonstrate the idea of conformal prediction using Japanese subnational age- and sex-specific mortality rates from 1975 to 2023, sourced from the \cite{JMD26}. The original data are registered (full count) mortality data \citep{DS41} and capture mortality rates for both sexes in the 47 prefectures, shown in Figure~\ref{fig:1}.

For analytical refinement, ages are grouped into single years from 0 to 94, with a final open-ended group of 95+. This treatment helps generate a comprehensive overview of mortality trends while bypassing noisy data associated with extremely high ages \citep[see also][]{HBY13}. 

Because of measurement errors, observed subnational mortality rates are noisy across ages. We employ penalized regression spline smoothing with a monotonic constraint above age 65 \citep{Wood94} to generate smooth functions and to handle missing data and values outside $[0,1]$. In functional data analysis, it is often assumed that the $\log$ mortality rate in each year follows an underlying smooth function of age, $\Y_t(u)$, such that
\begin{equation*}
\log\Z_{t}(u_{j}) = \log\Y_{t}(u_{j}) + \sigma_{t}(u_{j})\epsilon_{t,j},
\end{equation*} 
where $\Z_{t}(u_j)$ denotes the original mortality rate, and $\Y_t(u_j)$ denotes the smoothed mortality rate for $t=1,\dots, T$, $j=1,\dots,J$. The error term $\bm{\epsilon} = (\bm{\epsilon}_1,\dots,\bm{\epsilon}_J), \bm{\epsilon}_{j}=(\epsilon_{1,j},\dots,\epsilon_{T,j})^{\top}$ consists of independent and identically distributed variables with zero mean and unit variance, where $J$ denotes the number of ages, and $\sigma_t(u_j)$ allows for heteroskedasticity. The smoothness of age-specific mortality rates over age in each year is a distinguishing feature, which can improve short-term forecast accuracy \citep[see, e.g.,][]{BC19, YSR24}. Computationally, the \verb|smooth.demogdata| function in the demography package \citep{Hyndman25} is used to smooth the raw data.

\begin{figure}[!htb]
\centering
\includegraphics[width=8.7cm]{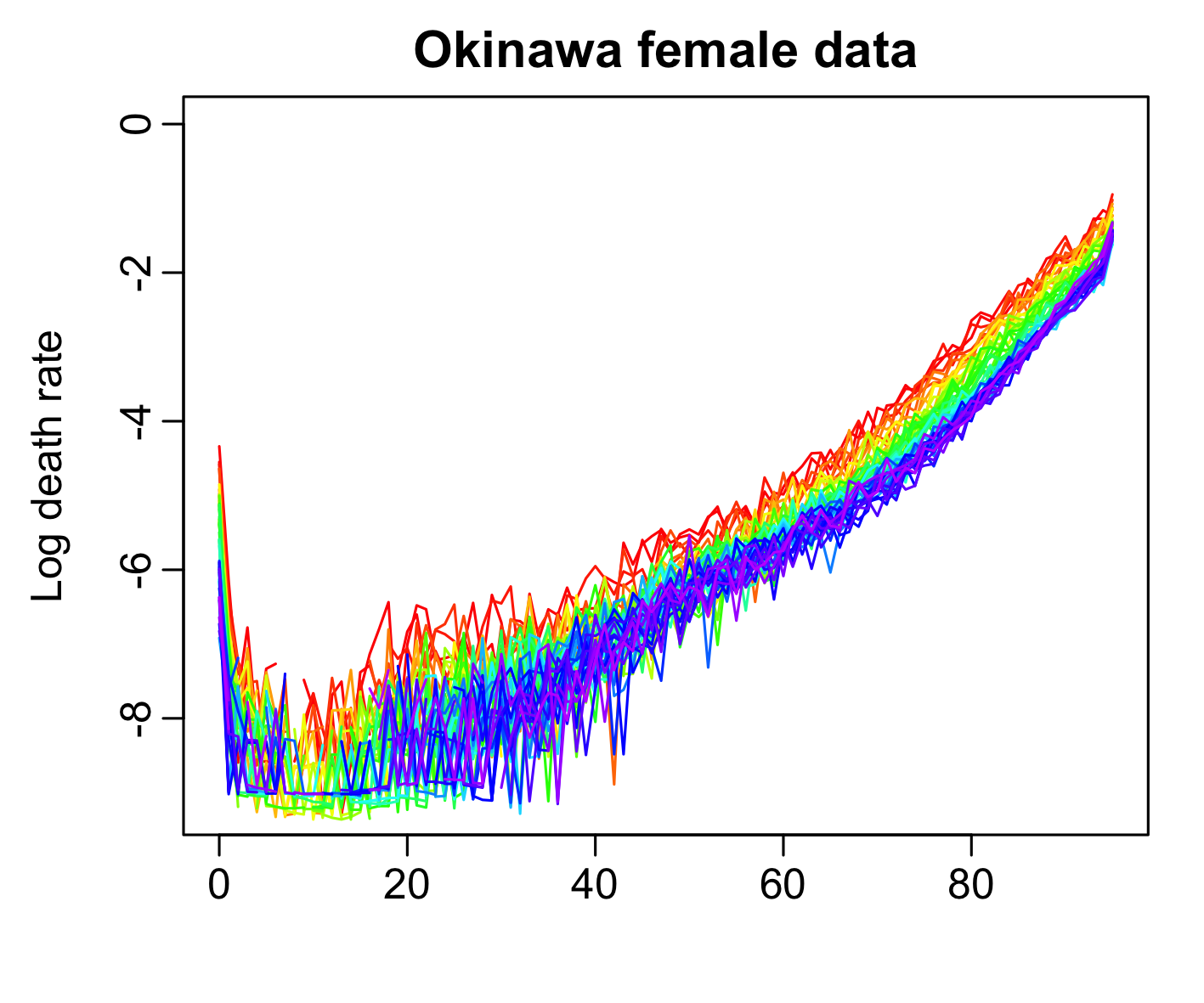}
\quad
\includegraphics[width=8.7cm]{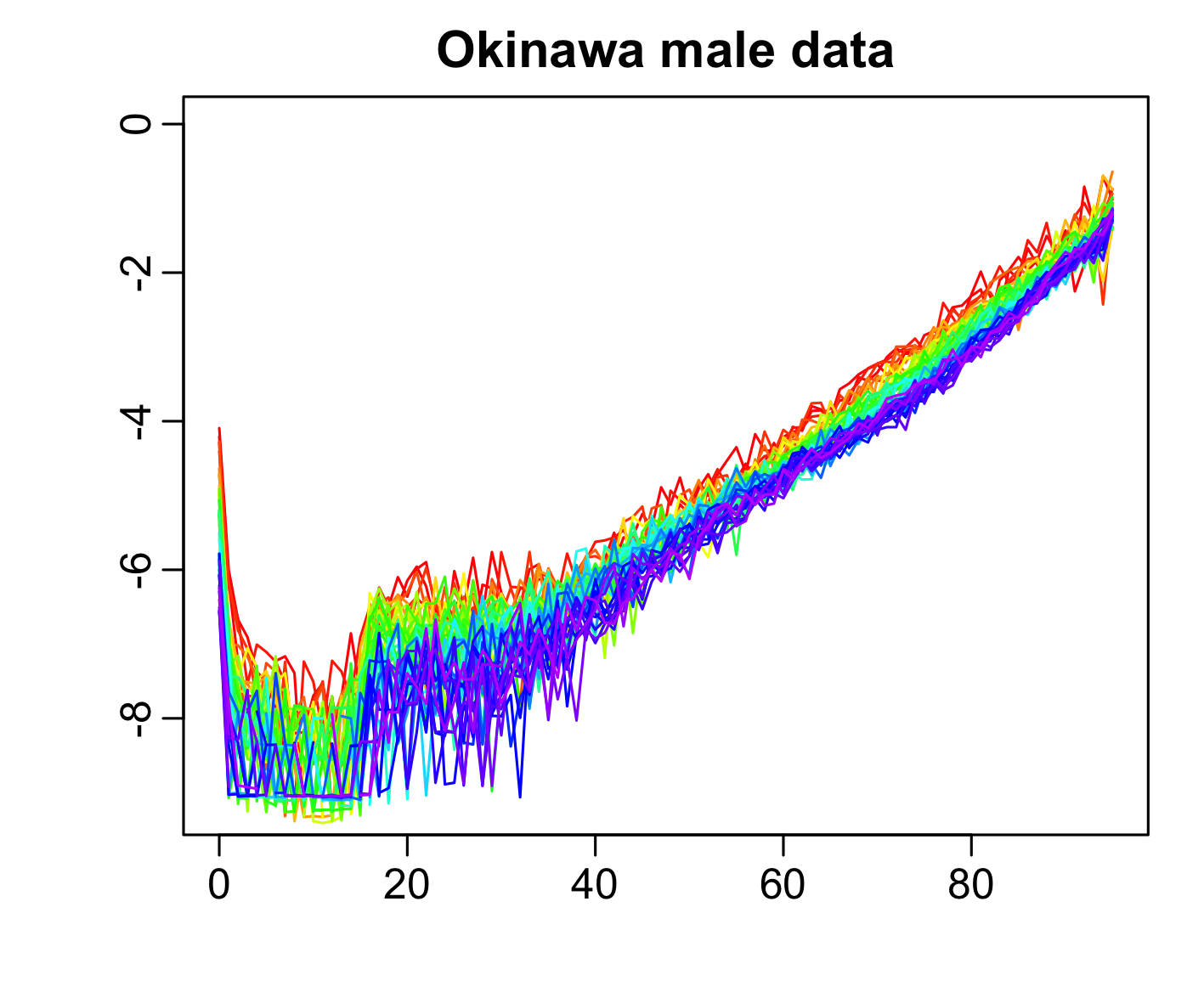}
\\
\includegraphics[width=8.7cm]{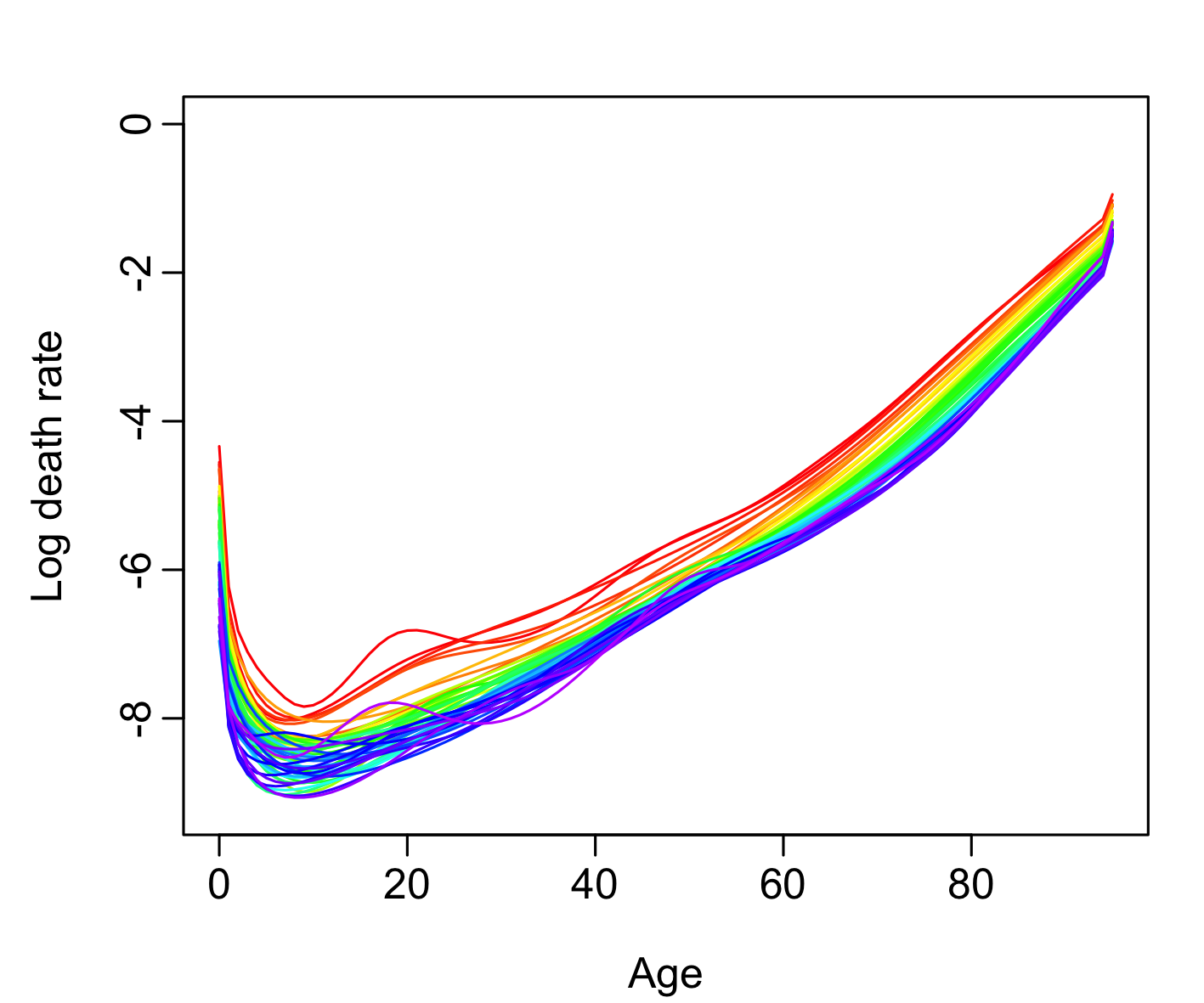}
\quad
\includegraphics[width=8.7cm]{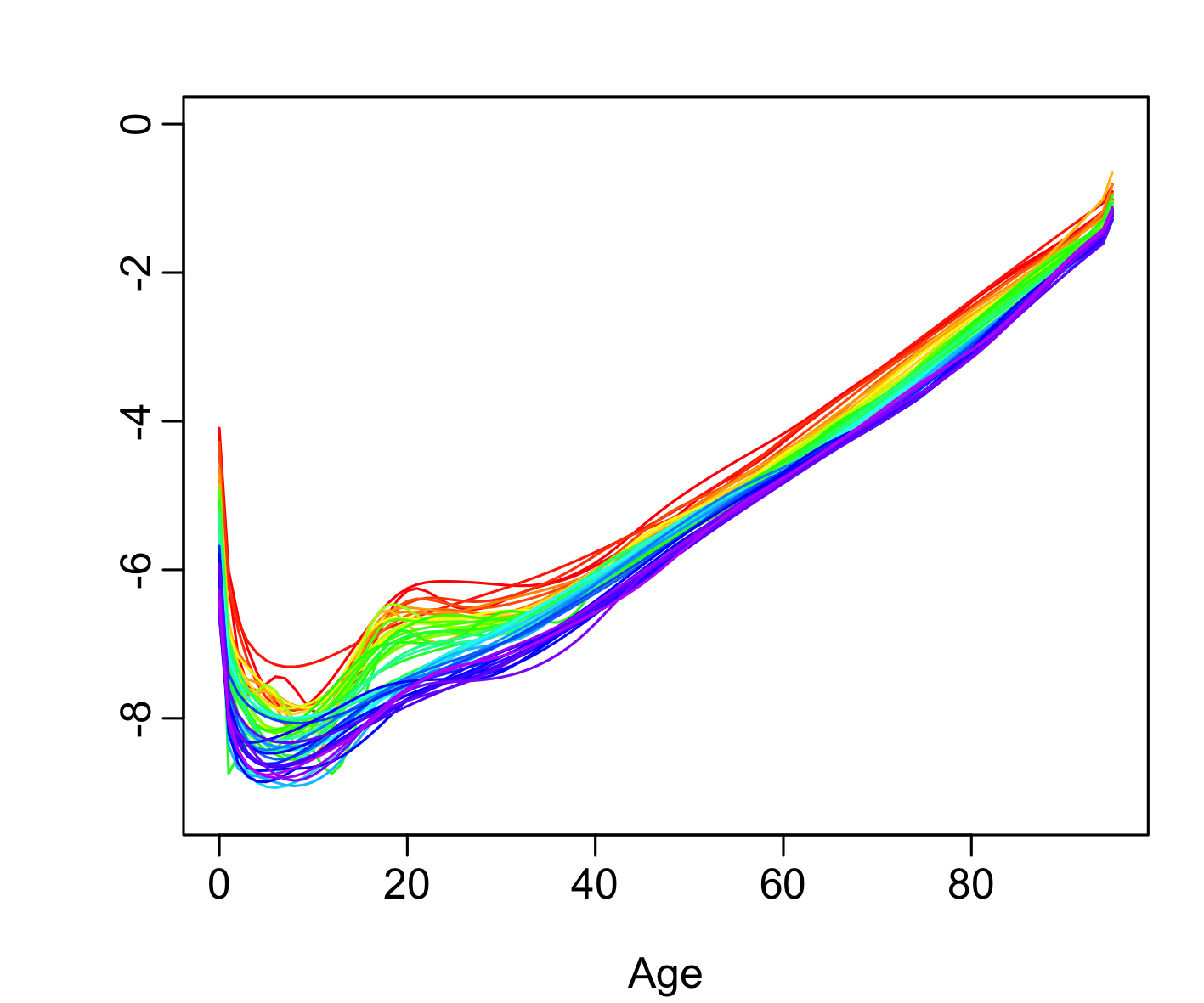}
\caption{\small The raw (first row) and smoothed (second row) age-specific $\log$ mortality rates between 1975 and 2023 in Okinawa, Japan. Curves are ordered chronologically using a rainbow color palette; the oldest curves are shown in red, and the most recent in violet.}\label{fig:2}
\end{figure}

To visualize the time trend, we present the age-specific mortality rates using a rainbow plot of \cite{HS10} in Figure~\ref{fig:2}. The $\log$ mortality rates from the distant past years are shown in red, while data from the most recent years are shown in violet. Figure~\ref{fig:2} shows typical mortality curves for a developed country, with rapidly decreasing mortality rates in the early years of life, followed by an increase during the teenage years, a plateau in young adulthood, and then a steady increase from about age 30 \citep[see also][]{JSS24}. By examining changes in mortality rates by age, year, and region, it becomes evident that mortality rates have undergone substantial variation over time. As in many developed nations, females in Japan have lower mortality rates than males at all ages. Because of these systematic biological differences,  we model the two sexes separately.

\section{Joint modeling of high-dimensional functional time series}\label{sec:3}

Since all subnational series are correlated in space and time, we present the method of \cite{SJ26} for modeling and forecasting HDFTS. It consists of a decomposition via one-way functional analysis of variance in Section~\ref{sec:3.1}, whose residuals are modeled via a functional factor model in Section~\ref{sec:3.2}. By forecasting the estimated factors, we obtain the forecasts of the residuals in Section~\ref{sec:3.3}. Section~\ref{sec:3.4} documents all steps in a computational algorithm.

\subsection{Decomposition via one-way functional analysis of variance}\label{sec:3.1}

As we observe multiple functional time series (i.e., subnational $\log$ age-specific mortality rates), we are interested in examining the effect at each prefecture, also known as the functional row effect. We denote the $N\times T$ array
\begin{equation*}
\begin{bmatrix}
\Y_{1,1}(u) & \Y_{2,1}(u) & \cdots & \Y_{T,1}(u) \\
\Y_{1,2}(u) & \Y_{2,2}(u) & \cdots & \Y_{T,2}(u) \\
\vdots & \vdots & \ddots & \vdots \\
\Y_{1,N}(u) & \Y_{2,N}(u) & \cdots & \Y_{T,N}(u)
\end{bmatrix}.
\end{equation*}

We consider an exact decomposition, known as the one-way functional ANOVA \citep[see, e.g.,][p.538]{Smaga26, CF10}. The $\log$ smoothed mortality rates can be decomposed as
\begin{equation}
\log\Y_{t,s}(u) = \underbrace{\theta(u)+\delta_s(u)}_{\text{deterministic}} + \underbrace{\X_{t,s}(u)}_{\text{time-varying}},\label{eq:FANOVA}
\end{equation}
where $\theta(u)$ represents a functional grand effect, $\delta_s(u)$ denotes the $s$\textsuperscript{th} functional row effect, and $\X_{t,s}(u)$ denotes the time-varying residual process. To estimate $\theta(u)$, $\delta_{s}(u)$ and $\X_{t,s}(u)$, we consider the functional median polish of \cite{SG12}, due to its robustness to outliers. 

The functional median polish extends the classical median polish procedure of \cite{EH83} to the functional data setting and retains robustness to outliers. As outlined in \cite{SG12}, the functional grand effect and row effects can be obtained through the following iterative procedure:
\begin{enumerate}
\item[1)] Compute the functional median for each row and record these values. Subtract the corresponding row functional median from each function within that row.
\item[2)] Compute the functional median of the row functional medians and denote it as the functional grand effect. Subtract this grand effect from each row's functional median, and record the resulting quantities as the functional row effects.
\item[3)] Repeat Steps 1) and 2), updating and accumulating the functional grand and row effects at each iteration, until the row functional medians stabilize and no further changes occur.
\end{enumerate}
Alternatively, one could also use the mean-based functional ANOVA as in \cite{SJ26}, which does not require any iterations. Computationally, it can be achieved via the \verb|One_way_mean| function in the hdftsa package \citep{Shang26} in~\Rlogo.

\subsection{A functional factor model}\label{sec:3.2}

In the HDFTS literature, there are several variants of the factor model:
\begin{inparaenum}[1)]
\item real-valued factor loadings and functional factors \citep{GQW+26}; 
\item functional factor loadings and real-valued factors \citep{TNH23};
\item functional factor loadings and functional factors \citep{LLS+26}.
\end{inparaenum}

For each sex, we consider the following functional factor model:
\begin{equation}
\X_{t,s}(u) = \sum_{k=1}^{k_{*}}\int_{\mathcal{C}_k^*}B_{k,s}(u,v)F_{t,k}(v)dv+\varepsilon_{t,s}(u).\label{eq:2}
\end{equation}
By imposing a low-dimensional functional factor condition on the latent factor $F_{t,k}(v)$, we obtain the series approximation
\begin{equation}
F_{t,k}(v) = \Phi_k(v)^{\top}G_t + \zeta_{t,k}(v),\quad v\in \mathcal{C}_k^*,\quad k=1,\dots,k_{*}, \label{eq:3}
\end{equation}
where $\Phi_k(\cdot)$ is a $q$-dimensional vector of basis functions, $G_t$ is a $q$-dimensional vector of stationary random variables, and $\zeta_{t,k}(v)$ is the model error.

By plugging~\eqref{eq:3} into~\eqref{eq:2}, we obtain
\begin{align}
\X_{t,s}(u) &= \sum_{k=1}^{k_{*}}\int_{\mathcal{C}_k^*}B_{k,s}(u,v)\big[\Phi_k(v)^{\top}G_t + \zeta_{t,k}(v)\big]dv+\varepsilon_{t,s}(u) \label{eq:4}\\
&=\sum_{k=1}^{k_{*}}\int_{\mathcal{C}_k^*}B_{k,s}(u,v)\Phi_k(v)^{\top}dvG_t+\sum_{k=1}^{k_{*}}\int_{\mathcal{C}_k^*}B_{k,s}(u,v)\zeta_{t,k}(v)dv + \varepsilon_{t,s}(u) \notag\\
&=\Lambda_s(u)^{\top}G_t + \varepsilon_{t,s}^*(u),\notag
\end{align}
where $\Lambda_s(u) = \sum_{k=1}^{k_{*}}\int_{\mathcal{C}_k^*}B_{k,s}(u,v)\Phi_k(v)^{\top}dv$ and $\varepsilon^*_{t,s}(u) = \sum_{k=1}^{k_{*}}\int_{\mathcal{C}_k^*}B_{k,s}(u,v)\zeta_{t,k}(v)dv + \varepsilon_{t,s}(u)$.

To estimate $G_t$, we begin with the estimation of the covariance of $\X_{t,s}(u)$ by
\begin{equation*}
\bm{\Delta} = (\Delta_{tt^{'}})_{T\times T}\quad \text{with}\quad \Delta_{tt^{'}} = \frac{1}{N}\sum^{N}_{s=1}\int_{u\in \mathcal{C}_s} \X_{t,s}(u)\X_{t^{'},s}(u)du.
\end{equation*}
Via the eigenanalysis of the $T\times T$ matrix $\bm{\Delta}$, we obtain $\widetilde{\bm{G}}=(\widetilde{G}_{1},\dots,\widetilde{G}_T)^{\top}$ as a $T\times q$ matrix with columns being the eigenvectors (multiplied by $\sqrt{T}$) corresponding to the $q$ largest estimated eigenvalues of~$\bm{\Delta}$, where $q$ is a proxy for $k_{\ast}$ in~\eqref{eq:4}. The factor loading functions are estimated as
\begin{equation*}
\big(\widetilde{\Lambda}_s(u): u\in \mathcal{C}_s\big) = \frac{1}{T}\sum^T_{t=1}\X_{t,s}(u)\widetilde{G}_t,\quad s=1,\dots,N,
\end{equation*}
via least squares, using the normalization restriction $\frac{1}{T}\sum^T_{t=1}\widetilde{G}_t\widetilde{G}_t^{\top} = \bm{I}_q$, where $\bm{I}_q$ is the  $q\times q$ identity matrix.

We consider an information criterion, introduced in \cite{LLS+26}, to estimate the factor number $q$, which can slowly diverge to infinity. Let $\nu_{\ell}(\bm{\Delta}/T)$ be the $\ell$\textsuperscript{th} largest eigenvalue of $\bm{\Delta}/T$ and define
\begin{equation*}
\widehat{q} = \argmin_{1\leq \ell\leq q_{\max}}\left[\nu_{\ell}(\bm{\Delta}/T)+\ell\cdot \phi_{N,T}\right]-1,
\end{equation*}
where $\phi_{N,T}$ is the penalty parameter and $q_{\max}$ is a user-specified positive integer. In practice, we set $\phi_{N,T} = [\max(T, N)]^{-\frac{1}{2}}$ and $q_{\max} = T$.

\subsection{Forecasting HDFTS}\label{sec:3.3}

Conditional on the estimated factor loadings $\widetilde{\Lambda}_s(u)$, the time-varying dynamics are captured by the $q$-dimensional factors $\widetilde{\bm{G}}_t$. By applying a univariate time-series forecasting method, such as exponential smoothing (ETS) and autoregressive integrated moving average (ARIMA), we obtain the $h$-step-ahead forecast $\widehat{\widetilde{\bm{G}}}_{T+h|T}$. The ETS and ARIMA models are selected automatically, based on the corrected Akaike information criterion (AIC$_{\text{c}}$) of \cite{HT89}, and automatic order selection algorithm implemented via \verb|auto.arima| function in the forecast package \citep{HK08} in~\Rlogo, respectively.

Multiplying the forecast factors by the estimated factor loadings gives the $h$-step-ahead forecast of $\X_{T+h,s}(u)$:
\begin{equation*}
\widehat{\X}_{T+h|T, s}(u) = \widetilde{\Lambda}_s(u)^{\top}\widehat{\widetilde{\bm{G}}}_{T+h|T},
\end{equation*}
where $\widehat{\widetilde{\bm{G}}}_{T+h|T}$ denotes the $h$-step-ahead time-series forecast of $\widetilde{\bm{G}}_{T+h}$.

By adding the deterministic mean in~\eqref{eq:FANOVA}, we obtain the forecast of $\log$ smooth mortality rates:
\begin{equation*}
\log\widehat{\Y}_{T+h|T,s}(u) = \theta(u) + \delta_s(u)+\widehat{\X}_{T+h|T, s}(u).
\end{equation*}

\subsection{Computational algorithm}\label{sec:3.4}

Our procedure consists of three stages. 
\begin{asparaenum}[1)]
\item In Stage~1 (Section~\ref{sec:3.1}), we apply the deterministic decomposition to separate the functional grand and row effects from the time-varying residuals. 
\item In Stage~2 (Section~\ref{sec:3.2}), we fit a functional factor model to those residuals. 
\item In Stage~3 (Section~\ref{sec:3.3}), we forecast the real-valued factors with a univariate time-series method and reassemble the forecast. Algorithm~\ref{alg:owfanova-ffm} summarizes the complete procedure.
\end{asparaenum}

\begin{algorithm}[!htb]
\caption{\small One-way functional ANOVA with a functional factor model for forecasting HDFTS}\label{alg:owfanova-ffm}
\begin{algorithmic}[1]
\Require Observed curves $\left\{\log\Y_{t,s}(u): t=1,\dots,T;\ s=1,\dots,N\right\}$ for one sex; forecast horizon $h$; upper bound $q_{\max}$; penalty $\phi_{N,T}$
\Ensure  $h$-step-ahead forecast $\log\widehat{\Y}_{T+h|T,s}(u)$ for each $s$

\Statex \textit{Stage 1: deterministic decomposition (Section~\ref{sec:3.1})}
\State Initialize $\widehat{\theta}(u)\gets 0$ and $\widehat{\delta}_s(u)\gets 0$, $s=1,\dots,N$; set $R_{t,s}(u)\gets\log\Y_{t,s}(u)$
\Repeat
  \State For each row $s$, compute the functional median $m_s(u)$ of $\{R_{t,s}(u)\}_{t=1}^{T}$ and set $R_{t,s}(u)\gets R_{t,s}(u)-m_s(u)$
  \State Compute $g(u)$, the functional median of $\{m_s(u)\}_{s=1}^{N}$, and set $m_s(u)\gets m_s(u)-g(u)$
  \State Accumulate $\widehat{\theta}(u)\gets\widehat{\theta}(u)+g(u)$ and $\widehat{\delta}_s(u)\gets\widehat{\delta}_s(u)+m_s(u)$
\Until{the row functional medians stabilize}
\State Set the residual curves $\X_{t,s}(u)\gets\log\Y_{t,s}(u)-\widehat{\theta}(u)-\widehat{\delta}_s(u)$
\Statex \textit{Stage 2: functional factor model (Section~\ref{sec:3.2})}
\State Form the $T\times T$ covariance matrix $\bm{\Delta}=(\Delta_{tt'})$ with $\Delta_{tt'}=N^{-1}\sum_{s=1}^{N}\int_{\mathcal{C}_s}\X_{t,s}(u)\X_{t',s}(u)\,\mathrm{d}u$
\State Select the factor number $\widehat{q}=\argmin_{1\leq\ell\leq q_{\max}}\bigl[\nu_{\ell}(\bm{\Delta}/T)+\ell\cdot\phi_{N,T}\bigr]-1$
\State Set $\widetilde{\bm{G}}=(\widetilde{G}_1,\dots,\widetilde{G}_T)^{\top}$ to $\sqrt{T}$ times the eigenvectors of $\bm{\Delta}$ associated with its
$\widehat{q}$ largest eigenvalues, so that $T^{-1}\sum_{t=1}^{T}\widetilde{G}_t\widetilde{G}_t^{\top}=\bm{I}_{\widehat{q}}$
\State Estimate the loading functions by least squares, $\widetilde{\Lambda}_s(u)=T^{-1}\sum_{t=1}^{T}\X_{t,s}(u)\widetilde{G}_t$

\Statex \textit{Stage 3: forecasting (Section~\ref{sec:3.3})}
\State Fit a univariate time-series model to each of the $\widehat{q}$ factor series, using either an ETS or an autoregressive integrated moving average (ARIMA) model, with the order selected by the corrected AIC$_{\text{c}}$, and compute $\widehat{\widetilde{\bm{G}}}_{T+h|T}$
\State Set $\widehat{\X}_{T+h|T,s}(u)= \widetilde{\Lambda}_s(u)^{\top}\widehat{\widetilde{\bm{G}}}_{T+h|T}$
\State \Return $\log\widehat{\Y}_{T+h|T,s}(u)=\widehat{\theta}(u)+\widehat{\delta}_s(u)+\widehat{\X}_{T+h|T,s}(u)$
\end{algorithmic}
\end{algorithm}

\section{Conformal prediction}\label{sec:4}

In the machine-learning field, the conformal prediction of \cite{SV08} is used to construct probabilistic forecasts calibrated on out-of-sample errors. Since its introduction by \cite{GVV98}, many variants have been proposed \citep[see][for a comprehensive overview]{AB23}. Among them, we focus on conformal prediction methods suitable for time series.

\subsection{Split conformal prediction}\label{sec:4.1}

As shown in Figure~\ref{fig:3}, we partition the 49-year sample (1975–2023) into training, validation, and test sets with proportions of approximately 60\%, 20\%, and 20\%, respectively. Using the initial training period from 1975 to 2002, we adopt an expanding-window forecasting scheme to generate $h$-step-ahead forecasts for the validation period 2003–2013, for $h = 1, 2, \dots, 10$ \citep[see also][for a functional time series]{SH26}. Under this scheme, the training sample is progressively enlarged at each iteration. The number of available curves in the validation or test set varies with the forecast horizon $h$: for instance, when $h = 1$, there are 11 years to compute residual functions, corresponding to the differences between the observed curves in the validation set and their forecasts; whereas when $h = 10$, only two years are available. For a given horizon $h$, denote the residual functions on the original scale by $\widehat{\varepsilon}_{m,s}(u) = \Z_{m,s}(u) - \widehat{\Z}_{m,s}(u)$ for $m = 1, \dots, M$, where $M$ is the number of years in the validation set. Following \citet{ANH15}, from these residual functions, we compute summary measures $\gamma_s(u)$ for each prefecture $s$, such as the pointwise standard deviation. Our objective is to determine a tuning parameter $\xi_{\alpha}$ such that $100(1-\alpha)\%$ of the residuals satisfy
\begin{equation*}
-\xi_{\alpha}\gamma_s(u)\leq \widehat{\bm{\varepsilon}}_{s}(u)\leq \xi_{\alpha}\gamma_s(u),
\end{equation*}
where $\widehat{\bm{\varepsilon}}_s(u) = [\widehat{\varepsilon}_{1,s}(u), \dots, \widehat{\varepsilon}_{M,s}(u)]$. By the law of large numbers, when $M$ is reasonably large, one obtains
\begin{equation*}
\text{Pr}\left[-\xi_{\alpha}\gamma_s(u)\leq \Z_{T+h,s}(u) - \widehat{\Z}_{T+h,s}(u)\leq \xi_{\alpha}\gamma_s(u)\right]\approx \frac{1}{M}\sum^M_{m=1}\mathds{1}\left\{-\xi_{\alpha}\gamma_s(u)\leq \widehat{\varepsilon}_{m,s}(u)\leq \xi_{\alpha}\gamma_s(u)\right\},
\end{equation*}
where $\Z_{T+h,s}(u)$ denotes the holdout age-specific mortality rates at time $T+h$ in prefecture~$s$, $\widehat{\Z}_{T+h,s}(u)$ denotes the corresponding forecasts on the original scale (after exponential back-transformation), and $\mathds{1}\{\cdot\}$ represents the binary indicator function. Note that for a small value of $M$, the split conformal prediction can under-cover relative to the nominal level.

\subsection{Non-conformity score conformal prediction}\label{sec:4.2}

Non-conformity score prediction interval uses the same data splitting scheme as the split conformal prediction, as shown in Figure~\ref{fig:3}: the point forecasts are obtained from an expanding window, and the interval is calibrated on the forecasts for the validation set (2003--2013) and evaluated on the forecasts for the test set (2014--2023). However, instead of multiplying the pointwise standard deviation $\gamma_s(u)$ by a tuned coefficient $\xi_\alpha$, it uses the width-standardized non-conformity score \citep{RPC19, KTL23}, and the prediction interval is set by an empirical quantile of these scores, so that no tuning is needed.
\begin{figure}[!htb]
\begin{center}
\begin{tikzpicture}[x=0.25cm, y=1cm, font=\footnotesize]
\draw[fill=blue!8] (0,0) rectangle (28,0.7) node[midway, align=center] {Training\\1975--2002};
\draw[fill=orange!35] (28,0) rectangle (39,0.7) node[midway, align=center] {Validation\\2003--2013};
\draw[fill=red!25] (39,0) rectangle (49,0.7) node[midway, align=center] {Test\\2014--2023};
\draw[fill=green!30] (17,0.7) rectangle (28,1.4) node[midway, align=center] {Calibration\\1992--2002};
\foreach \xx in {17,28,39} \draw[dashed, gray] (\xx,0) -- (\xx,-6.35);
\foreach \o/\yy/\col in {
1991/-0.55/green!30, 
1992/-1.0/green!30, 
2001/-1.9/green!30,
2002/-2.55/orange!35, 
2003/-3.0/orange!35, 
2012/-3.9/orange!35,
2013/-4.55/red!25, 
2014/-5.0/red!25, 
2022/-5.9/red!25}{
  \pgfmathsetmacro{\e}{\o-1974}
  \draw[fill=blue!20] (0,{\yy-0.17}) rectangle (\e,{\yy+0.17}) node[midway] {Train};
  \draw[fill=\col] (\e,{\yy-0.17}) rectangle ({\e+1},{\yy+0.17}) node[midway] {\scriptsize F};
  \node[left] at (0,\yy) {1975:\o};
}
\foreach \yy in {-1.45,-3.45,-5.45} \node[left] at (0,\yy) {$\vdots$};
\draw[->] (0,-6.35) -- (52,-6.35) node[right] {Time};
\foreach \yr in {1975,1992,2003,2014,2023}{
  \pgfmathsetmacro{\c}{\yr-1974.5}
  \draw (\c,-6.35) -- (\c,-6.45) node[below] {\scriptsize\yr};
}
\draw[fill=blue!20] (31,1.15) rectangle (33,1.4);
\node[right] at (33,1.275) {Training window};
\draw[fill=green!30] (31,0.8) rectangle (31.667,1.05);
\draw[fill=orange!35] (31.667,0.8) rectangle (32.333,1.05);
\draw[fill=red!25] (32.333,0.8) rectangle (33,1.05);
\node[right] at (33,0.925) {Forecast (F) when $h=1$};
\end{tikzpicture}
\end{center}
\caption{\small Data split scheme, shown for $h=1$. Each row is one expanding-window forecast, with its forecast (F) colored by the set it falls in. The calibration set (1992--2002) is used by adaptive conformal prediction only; for the split and non-conformity score predictions, these years are part of the training set. The validation and test sets are shared by all three methods.}\label{fig:3}
\end{figure}

The method first needs an initial interval, which gives the shape of the prediction interval across ages in which the scores are measured. Based on each validation forecast, we take the initial interval $\widehat{\mathcal Z}_{m,s}(u)\pm\gamma_s(u)$ of interval width $2\gamma_s(u)$, where $\gamma_s(u)$ is the same pointwise standard deviation of the validation residuals as in Section~\ref{sec:4.1}. The non-conformity score of an observation measures the distance outside the initial interval, standardized in widths of the initial interval,
\begin{equation}
S_{m,s}(u)=\frac{\max\big\{\left[\widehat{\mathcal Z}_{m,s}(u)-\gamma_s(u)\big]-\mathcal Z_{m,s}(u),\;\mathcal Z_{m,s}(u)-\big[\widehat{\mathcal Z}_{m,s}(u)+\gamma_s(u)\right]\big\}}{2\gamma_s(u)},
\label{eq:ns-score}
\end{equation}
which is negative inside the initial interval and positive outside. Because the scores are standardized, they are comparable across ages. For each prefecture and horizon, we pool the $n = MJ$ scores of the $M$ validation years and $J$ ages. Let $\widehat Q_s(\alpha)$ be the $k$-th smallest value, with $k=\lceil(n+1)(1-\alpha)\rceil$. The non-conformity score prediction interval is obtained by shifting each bound of the initial interval by $\widehat Q_s(\alpha)$ initial-interval widths, that is, by $2\widehat Q_s(\alpha)\gamma_s(u)$:
\begin{equation}
\begin{split}
&\left[\,\widehat{\mathcal Z}_{T+h,s}(u)-\gamma_s(u)-2\widehat Q_s(\alpha)\gamma_s(u),\;\widehat{\mathcal Z}_{T+h,s}(u)+\gamma_s(u)+2\widehat Q_s(\alpha)\gamma_s(u)\,\right]\\
&\qquad=\widehat{\mathcal Z}_{T+h,s}(u)\pm\left(1+2\widehat Q_s(\alpha)\right)\gamma_s(u).
\end{split}
\label{eq:ns-interval}
\end{equation}

The non-conformity score prediction interval has the same form as the split conformal interval, with $\xi_\alpha=1+2\widehat Q_s(\alpha)$, which is the smallest multiplier that keeps $k$ of the $n$ validation residuals inside the interval. The two methods thus differ only in how the multiplier is chosen: the split conformal prediction tunes $\xi_\alpha$ so that about $100(1-\alpha)\%$ of the validation residuals lie inside, whereas the non-conformity score prediction takes it from the quantile of the standardized non-conformity scores. The performances of the two intervals are expected to be close at short horizons, where many validation residuals are available. At longer horizons, however, they can differ in practice. In both methods, the validation coverage is a step function of the multiplier. With few validation residuals, the largest standardized residuals are few and far apart, and the coverage stays flat over wide ranges of the multiplier. Split conformal prediction finds $\xi_\alpha$ by a numerical search that compares the coverage at candidate $\xi_\alpha$ values, so it can stop on such a flat objective function, different from the required value. In contrast, the non-conformity score prediction interval reads its multiplier off directly as the $k$-th smallest score, without any tuning or searching, so it is not affected by the flat objective function and is computationally more efficient.

\subsection{Sequential predictive conformal prediction}\label{sec:4.3}

Without the need for a validation set, sequential predictive conformal prediction can automatically tune the predictive quantiles of the absolute residual functions as new data arrive. For instance, with the last 10 years as the test set, we use the other years to compute the absolute residuals $\{|\widehat{\varepsilon}_{3,s}(u_j)|,\dots,|\widehat{\varepsilon}_{\iota,s}(u_j)|\}$ for a given age $u_j$. We require at least the first two curves to produce a forecast, so residuals begin with the third curve. 

At the $1-\alpha$ quantile level, we fit a quantile regression on lagged residuals, where the order of autoregression, denoted by $p$, is determined by an information criterion, such as the AIC. Conditional on the $p$ most recent absolute residuals, we predict a one-step-ahead quantile, denoted as $\widehat{q}_{\iota+1, s}^{\alpha}(u_j)$, where $\iota$ denotes the index of the final observation in the training period. The prediction intervals are then given by 
\begin{equation*}
\widehat{\Z}_{\iota+1,s}(u_j)\pm \widehat{q}_{\iota+1, s}^{\alpha}(u_j), \qquad j=1,\dots,J.
\end{equation*} 
Once the actual curve $\Z_{\iota+1, s}(u)$ arrives, we can update the absolute residual $|\widehat{\varepsilon}_{\iota+1,s}(u_j)|$ and refit.

\subsection{Adaptive conformal prediction}\label{sec:4.4}

Adaptive conformal prediction, proposed by \citet{GC21}, starts from a base interval and adjusts it over time based on the performance of the past intervals. In this study, we take the non-conformity score prediction interval in \eqref{eq:ns-interval} as the base interval, but update the significance level over time based on the coverage of the past intervals: too many misses lower the level and widen the following prediction intervals, and vice versa. The interval for the forecast made in year~$T$ is
\begin{equation}
\widehat{\mathcal Z}_{T+h,s}(u)\pm\big[1+2\widehat Q_s(a_{T,s})\big]\gamma_s(u),
\label{eq:ac-interval}
\end{equation}
where $\widehat Q_s(a_{T,s})$ is the quantile of the scores at the significance level $a_{T,s}$ of year $T$. The level is updated once a year, when the observations of that year become available. In year $T$, the interval for this year, made in year $T-h$ at level $a_{T-h,s}$, can be checked, and the miscoverage proportion is
\begin{equation*}
\mathrm{err}_{T,s}=\frac{1}{J}\sum_{j=1}^{J}\mathds{1}\left\{\big|\mathcal Z_{T,s}(u_j)-\widehat{\mathcal Z}_{T,s}(u_j)\big|>\big[1+2\widehat Q_s(a_{T-h,s})\big]\gamma_s(u_j)\right\},
\end{equation*}
where $\widehat{\mathcal Z}_{T,s}(u_j)$ is the forecast for year $T$ made in year $T-h$. The significance level at time $T$ is then updated as
\begin{equation}
a_{T,s}=a_{T-1,s}+\lambda_s\,(\alpha-\mathrm{err}_{T,s}),
\label{eq:ac-update}
\end{equation}
where $\alpha$ is the nominal significance level and $\lambda_s\geq0$ is the learning rate, which is chosen to minimize the coverage probability difference over the $J$ ages and the validation years affected by~$\lambda_s$. The level $a_{T,s}$ thus reflects the coverage of all intervals checked up to year $T$.

The data splitting scheme of adaptive conformal prediction differs from that of the split and non-conformity score predictions (see Figure~\ref{fig:3}). In those two methods, the forecasts for the validation set (2003--2013) give the pointwise standard deviation $\gamma_s(u)$ and the non-conformity scores, and the resulting intervals are evaluated on the test set (2014--2023). Adaptive conformal prediction also has the additional parameter to choose, the learning rate $\lambda_s$, and therefore needs two sets before the test set: a calibration set, whose forecast errors give $\gamma_s(u)$ and the scores, and hence the construction of the base interval, and a validation set, on which the significance level is updated from the base interval as in~\eqref{eq:ac-update} and $\lambda_s$ is chosen from the coverage of the resulting intervals. For a consistent comparison, all methods calibrate their intervals on data of the same length and are evaluated on the same test set.

The calibration set consists of 11 years from 1992 to 2002, the same length as the validation set. Its forecasts, made in $1991,\dots,(2002-h)$, give $\gamma_s(u)$ and the non-conformity scores, and hence the base interval. Adaptive conformal prediction is then run on the forecasts for the validation set, made in $2002,\dots,(2013-h)$. The validation errors have the delayed feedback: an interval made in year $T$ is for year $T+h$, so its error is known only in year $T+h$, and the first error on the validation set, that of the interval made in 2002, is known in year $2002+h$. Hence, $a_{T,s}=\alpha$ for $T=2002,\dots,(2001+h)$ as no errors are yet known to update $a_{T,s}$, and $a_{T,s}$ follows~\eqref{eq:ac-update} for $T=(2002+h),\dots,2013$. As illustrated in Figure~\ref{fig:4}, when $h=5$, the intervals made in 2002--2006 use $\alpha$, and the significance level is first updated in 2007. The only intervals made with an updated significance level are thus those made in 2007 and 2008, for 2012 and 2013, so $\lambda_s$ is chosen to minimize the coverage probability difference averaged over these two years. Due to the limited length of the data, the 11-year validation set is thus too short to tune $\lambda_s$ for $h\geq6$, and we apply adaptive conformal prediction for $h=1,\dots,5$ only.

We tune $\lambda_s$ over $[0,0.15]$, which covers the learning rates used by \citet{GC24}, from 0.001 to 0.128. As illustrated in Figure~\ref{fig:4}, $\lambda_s$ is fixed after tuning, while the level $a_{T,s}$ keeps being updated through the validation and test sets. The validation run with the chosen $\lambda_s$ ends with $a_{2013,s}$, which includes the errors of all validation intervals. The adaptive conformal prediction continues from this level on the forecasts for the test set, made in $2013,\dots,(2023-h)$, whose intervals are evaluated. As on the validation set, the first test error is known in year $2013+h$, so $a_{T,s}=a_{2013,s}$ for $T=2013,\dots,(2012+h)$, and $a_{T,s}$ follows~\eqref{eq:ac-update} for $T=(2013+h),\dots,(2023-h)$; for $h=5$, the intervals made in 2013--2017 use $a_{2013,s}$, and the significance level is first updated in~2018.

\begin{figure}[!htb]
\centering
\resizebox{\linewidth}{!}{
\begin{tikzpicture}[x=0.62cm, y=1cm, >={Stealth[length=4pt]}, font=\footnotesize]
\draw[fill=orange!35] (0.5,0.75) rectangle (11.5,1.25) node[midway] {Validation 2003--2013: $\lambda_s$ chosen};
\draw[fill=red!25] (11.5,0.75) rectangle (21.5,1.25) node[midway] {Test 2014--2023: $\lambda_s$ fixed};
\draw (-0.3,0) -- (21.5,0);
\foreach \k in {0,...,21} 
{
\pgfmathtruncatemacro{\yr}{2002+\k}
\draw (\k,0) -- (\k,0.08) node[rotate=45, anchor=south west, inner sep=1pt] {\scriptsize\yr};
}
\draw[dashed, gray] (11.5,0) -- (11.5,-0.95);
\node[left] at (-0.3,-0.55) {\scriptsize level $a_{T,s}$};
\foreach \k in {0,...,4} \node[rotate=45, gray!70!black] at (\k,-0.55) {\scriptsize $\alpha$};
\foreach \k/\lab in {5/a_{2007,s}, 6/a_{2008,s}, 11/a_{2013,s}, 16/a_{2018,s}, 21/a_{2023,s}} \node[rotate=45, blue!70!black, fill=white, inner sep=1pt] at (\k,-0.55) {\scriptsize $\lab$};
\foreach \k in {12,...,15} \node[rotate=45, gray!70!black, fill=white, inner sep=1pt] at (\k,-0.55) {\scriptsize $a_{2013,s}$};
\foreach \k in {8.5,18.5} \node[rotate=45, blue!70!black] at (\k,-0.55) {$\cdots$};
\foreach \o/\yy/\lab/\col in {2002/-1.25/\alpha/gray!70!black, 2003/-1.7/\alpha/gray!70!black, 2004/-2.15/\alpha/gray!70!black, 2005/-2.6/\alpha/gray!70!black, 2006/-3.05/\alpha/gray!70!black, 2007/-3.5/a_{2007,s}/orange!85!black, 2008/-3.95/a_{2008,s}/orange!85!black, 2013/-4.95/a_{2013,s}/red!75!black, 2014/-5.4/a_{2013,s}/red!75!black, 2015/-5.85/a_{2013,s}/red!75!black, 2016/-6.3/a_{2013,s}/red!75!black, 2017/-6.75/a_{2013,s}/red!75!black, 2018/-7.2/a_{2018,s}/red!75!black}
{
  \draw[->, thick, \col] ({\o-2002},\yy) -- ({\o-1997},\yy);
  \draw[\col, fill=white] ({\o-2002},\yy) circle (2pt);
  \node[anchor=east, \col] at ({\o-2002.15},\yy) {$\lab$};
}
\foreach \xx/\yy in {8.5/-4.45, 18.5/-7.6} \node[gray!70!black] at (\xx,\yy) {$\ddots$};
\draw[->, dashed, blue!70!black] (5,-1.32) -- (5,-3.43) node[pos=0.5, left=-1pt, fill=white, inner sep=1pt] {$\mathrm{err}_{2007,s}$};
\draw[->, dashed, blue!70!black] (6,-1.77) -- (6,-3.88) node[pos=0.5, right=-1pt, fill=white, inner sep=1pt] {$\mathrm{err}_{2008,s}$};
\draw[->, dashed, blue!70!black] (16,-5.02) -- (16,-7.13) node[pos=0.5, right=-1pt, fill=white, inner sep=1pt] {$\mathrm{err}_{2018,s}$};
\draw[decorate, decoration={brace, amplitude=4pt}, orange!85!black, thick] (11.3,-3.38) -- (11.3,-4.07)
node[midway, right=6pt, align=left, black]{$\lambda_s$ minimizes the CPD of\\ these two intervals, over $[0,0.15]$};
\end{tikzpicture}
}
\caption{\small Updating of $a_{T,s}$ and choice of $\lambda_s$ for $h=5$ and a given prefecture $s$. Each row is one forecast, from the year it is made (circle) to the year it forecasts (arrowhead), labeled with the level its interval uses; the vertical dots mark years in which no forecast is made. The row under the axis gives $a_{T,s}$ in each year, in blue when it is updated, and the dashed arrows show three of the errors that update it. The orange rows are the two validation intervals that depend on $\lambda_s$.}\label{fig:4}
\end{figure}

\section{Evaluation of interval forecast accuracy}\label{sec:5}

\subsection{Expanding-window forecast scheme}\label{sec:5.1}

We partition the Japanese age-specific mortality data into a training period (1975–2002), a validation period (2003–2013), and a test period (2014–2023). The 10-year test sample enables us to assess forecast horizons from $h = 1$ to 10, reflecting varying levels of forecast uncertainty; for a shorter horizon, forecast uncertainty is comparatively smaller \citep[see also][]{SH26}. For some summary measures, such as quantiles or standard deviations, at least two years of observations are required for reliable computation at $h=10$. Thus, the validation period spans 11 years, instead of 10 years.

We adopt an expanding-window forecasting scheme, which is widely used in time-series analysis to evaluate model adequacy and parameter stability over time (see \citealp[Chapter 9]{ZW06}). For evaluating out-of-sample forecast accuracy, using the initial 39 years (1975–2013), we generate one- to ten-step-ahead forecasts. The training sample is then expanded to include 2014 (i.e., 1975–2014), the model parameters are re-estimated, and new one- to ten-step-ahead forecasts are produced. This procedure continues recursively, increasing the sample size by one year at each step until the end of the observation period in 2023. As a result, we obtain 10 one-step-ahead forecasts, 9 two-step-ahead forecasts, and so on, up to a single 10-step-ahead forecast. Figure~\ref{fig:5} illustrates the expanding-window scheme for the case $h = 1$, while similar procedures are applied for forecast horizons $h = 2, \dots, 10$.
\begin{figure}[!htb]
\begin{center}
\begin{tikzpicture}
\draw[->] (0,0) -- (10,0) node[right] {Time};
\draw[fill=blue!20] (0,-0.5) rectangle (3,0.5) node[midway] {Train};
\draw[fill=red!20] (3,-0.5) rectangle (3.5,0.5) node[midway] {F};
\draw[fill=blue!20] (0,-1.5) rectangle (5,-0.5) node[midway] {Train};
\draw[fill=red!20] (5,-1.5) rectangle (5.5,-0.5) node[midway] {F};
\draw[fill=blue!20] (0,-2.5) rectangle (7,-1.5) node[midway] {Train};
\draw[fill=red!20] (7,-2.5) rectangle (7.5,-1.5) node[midway] {F};
\draw[fill=blue!20] (0,-3.5) rectangle (9,-2.5) node[midway] {Train};
\draw[fill=red!20] (9,-3.5) rectangle (9.5,-2.5) node[midway] {F};
\node[left] at (0,0) {1975:2013};
\node[left] at (0,-1) {1975:2014};
\node[left] at (0,-2) {\hspace{-0.8in}{$\ddots$}};
\node[left] at (0,-3) {1975:2022};
\draw[fill=blue!20] (6.5,1) rectangle (7,1.5);
\node[right] at (7,1.25) {Training Window};
\draw[fill=red!20] (6.5,0.5) rectangle (7,1);
\node[right] at (7,0.75) {Forecast (F) when $h=1$};
\end{tikzpicture}
\end{center}
\caption{\small A diagram of the expanding-window forecast scheme, for the observations in the test set. The same scheme is also applied to calibrate tuning parameters using a validation set in the split, non-conformity score, and adaptive conformal prediction method. In the forecast accuracy evaluation, we consider the performances of methods at each forecast horizon.}\label{fig:5}
\end{figure}

\subsection{Measure of interval forecast accuracy}\label{sec:5.2}

To evaluate and compare interval forecast accuracy, we consider the empirical coverage probability (ECP), the coverage probability difference (CPD), and the interval score of \cite{GR07}. For each year in the test set, the $h$-step-ahead prediction intervals are calculated at the $1-\alpha$ nominal coverage probability; the lower and upper bounds, denoted by $[\widehat{\Z}^{\text{lb}}_{\eta+\xi,s}(u), \widehat{\Z}^{\text{ub}}_{\eta+\xi,s}(u)]$, need not be symmetric about the point forecasts \citep[see also][]{JSS24,SH26}. For a region $s$, the ECP and CPD are defined as
\begin{align*}
\text{ECP}_{\alpha,h,s} &= \frac{1}{101\times (11-h)}\sum^{10}_{\xi=h}\sum^{101}_{j=1}\mathds{1}\left\{\widehat{\Z}_{\eta+\xi,s}^{\text{lb}}(u_j)\leq \Z_{\eta+\xi,s}(u_j)\leq \widehat{\Z}_{\eta+\xi,s}^{\text{ub}}(u_j)\right\}, \\
\text{CPD}_{\alpha,h,s} &= \left|\frac{1}{101\times (11-h)}\sum^{10}_{\xi=h}\sum^{101}_{j=1}\left[\mathds{1}\{\Z_{\eta+\xi,s}(u_j)>\widehat{\Z}_{\eta+\xi,s}^{\text{ub}}(u_j)\}+\mathds{1}\{\Z_{\eta+\xi,s}(u_j)<\widehat{\Z}_{\eta+\xi,s}^{\text{lb}}(u_j)\}\right]-\alpha\right|,
\end{align*}
where $\eta$ represents the jump-off year, which is the base year from which forecasts are computed.

The ECP assesses coverage without evaluating the sharpness of the prediction interval. To combine coverage and sharpness, we use a scoring rule for the prediction interval at age $u_j$, defined~as
\begin{align*}
S_{\alpha,\xi}[\widehat{\Z}_{\eta+\xi,s}^{\text{lb}}(u_j), \widehat{\Z}_{\eta+\xi,s}^{\text{ub}}(u_j), \Z_{\eta+\xi, s}(u_j)] =\ & \Big[\widehat{\Z}_{\eta+\xi,s}^{\text{ub}}(u_j) - \widehat{\Z}_{\eta+\xi,s}^{\text{lb}}(u_j)\Big] \\
&+\frac{2}{\alpha}\Big[\widehat{\Z}_{\eta+\xi,s}^{\text{lb}}(u_j) - \Z_{\eta+\xi,s}(u_j)\Big]\mathds{1}\left\{\Z_{\eta+\xi,s}(u_j) <\widehat{\Z}_{\eta+\xi,s}^{\text{lb}}(u_j)\right\} \\
&+\frac{2}{\alpha}\Big[\Z_{\eta+\xi,s}(u_j) - \widehat{\Z}_{\eta+\xi,s}^{\text{ub}}(u_j)\Big]\mathds{1}\left\{\Z_{\eta+\xi,s}(u_j) > \widehat{\Z}_{\eta+\xi,s}^{\text{ub}}(u_j)\right\}.
\end{align*}
Hence, the ECP reported here measures age-averaged pointwise coverage rather than simultaneous coverage over the age domain.

Across ages and years in the test period, the MIS is given by
\begin{equation*}
\overline{S}_{\alpha,h,s} = \frac{1}{101\times (11-h)}\sum^{10}_{\xi=h}\sum^{101}_{j=1}S_{\alpha,\xi}\Big[\widehat{\Z}_{\eta+\xi,s}^{\text{lb}}(u_j), \widehat{\Z}_{\eta+\xi,s}^{\text{ub}}(u_j), \Z_{\eta+\xi,s}(u_j)\Big].
\end{equation*}
Given the same CPD, the MIS rewards narrower prediction intervals. It provides a second criterion for comparing interval forecast accuracy.

\section{Japanese subnational mortality forecasting results}\label{sec:6}

In Table~\ref{tab:1}, we present one- to ten-step-ahead interval forecast accuracy results for the conformal prediction methods at the nominal coverage probability of 95\%. Because of the equivalent forms discussed in Section \ref{sec:4.2}, split conformal and non-conformity score prediction intervals have very similar coverage at shorter horizons, where there are enough non-conformity scores. They differ mainly at longer horizons (e.g., $h=9$ and 10), where $\gamma_s(u)$ and the scores are obtained from only a few validation forecasts. In such cases, as explained in Section~\ref{sec:4.2}, the numerical search for $\xi_\alpha$ in split conformal prediction often stops different from the required value, whereas non-conformity score prediction takes the exact quantile of the scores, so its intervals become much wider than those of split conformal prediction. Additionally, both split and non-conformity score conformal prediction tend to \textit{under-cover}, with ECPs below the nominal level. In contrast, the sequential predictive conformal prediction tends to \textit{over-cover}.

\begin{center}
\tabcolsep 0.05in
\renewcommand{\arraystretch}{0.93}
\begin{small}
\begin{longtable}{@{}lllrrrrrrrrrrrr}
\caption{\small Averaged over 47 prefectures and ages in Japan, the one- to ten-step-ahead interval forecast accuracy is presented for the conformal prediction methods, namely split, sequential, non-conformity score, and adaptive conformal predictions, at the nominal coverage probability of 95\%. In split conformal prediction, the interval is based on the standard deviation (sd) of the validation residuals. The adaptive conformal prediction is computed for $h=1,\dots,5$ only; the mean over $h=1,\dots,5$ compares all four methods, and the mean over $h=1,\dots,10$ the other three. For forecasting the estimated factors, we consider two univariate time-series forecasting methods, namely ARIMA and ETS. In each row, the smallest CPD and the smallest MIS are shown in red and blue, respectively.}\label{tab:1}\\
\toprule
 & & & \multicolumn{3}{c}{Split (sd)} & \multicolumn{3}{c}{Sequential} & \multicolumn{3}{c}{\begin{tabular}[b]{@{}c@{}}Non-conformity\\ Score\end{tabular}} & \multicolumn{3}{c}{Adaptive} \\
\cmidrule(lr){4-6}\cmidrule(lr){7-9}\cmidrule(lr){10-12}\cmidrule(lr){13-15}
Method & Sex & $h$ & ECP & CPD & MIS & ECP & CPD & MIS & ECP & CPD & MIS & ECP & CPD & MIS \\
\midrule
\endfirsthead
\toprule
 & & & \multicolumn{3}{c}{Split (sd)} & \multicolumn{3}{c}{Sequential} & \multicolumn{3}{c}{\begin{tabular}[b]{@{}c@{}}Non-conformity\\ Score\end{tabular}} & \multicolumn{3}{c}{Adaptive} \\
\cmidrule(lr){4-6}\cmidrule(lr){7-9}\cmidrule(lr){10-12}\cmidrule(lr){13-15}
Method & Sex & $h$ & ECP & CPD & MIS & ECP & CPD & MIS & ECP & CPD & MIS & ECP & CPD & MIS \\
\midrule
\endhead
\midrule
\multicolumn{15}{r}{Continued on next page} \\
\endfoot
\endlastfoot
ARIMA & F & 1 & 0.934 & 0.022 & \textcolor{colMIS}{0.008} & 0.969 & 0.029 & 0.013 & 0.935 & 0.021 & \textcolor{colMIS}{0.008} & 0.955 & \textcolor{colCPD}{0.011} & 0.011 \\
 &  & 2 & 0.928 & 0.029 & \textcolor{colMIS}{0.009} & 0.971 & 0.031 & 0.014 & 0.929 & 0.028 & \textcolor{colMIS}{0.009} & 0.958 & \textcolor{colCPD}{0.013} & 0.012 \\
 &  & 3 & 0.919 & 0.035 & \textcolor{colMIS}{0.010} & 0.974 & 0.030 & 0.015 & 0.920 & 0.034 & \textcolor{colMIS}{0.010} & 0.956 & \textcolor{colCPD}{0.015} & 0.012 \\
 &  & 4 & 0.915 & 0.040 & \textcolor{colMIS}{0.011} & 0.975 & 0.031 & 0.016 & 0.916 & 0.039 & \textcolor{colMIS}{0.011} & 0.956 & \textcolor{colCPD}{0.015} & 0.013 \\
 &  & 5 & 0.901 & 0.052 & 0.014 & 0.976 & 0.032 & 0.018 & 0.904 & 0.050 & \textcolor{colMIS}{0.013} & 0.954 & \textcolor{colCPD}{0.018} & 0.014 \\
\cmidrule{3-15}
 & & Mean (1--5) & 0.919 & 0.036 & \textcolor{colMIS}{0.010} & 0.973 & 0.031 & 0.015 & 0.921 & 0.034 & \textcolor{colMIS}{0.010} & 0.956 & \textcolor{colCPD}{0.014} & 0.012 \\
\cmidrule{3-15}
 &  & 6 & 0.896 & 0.057 & 0.017 & 0.974 & \textcolor{colCPD}{0.032} & 0.019 & 0.900 & 0.054 & \textcolor{colMIS}{0.016} &  &  &  \\
 &  & 7 & 0.903 & 0.052 & 0.019 & 0.971 & \textcolor{colCPD}{0.031} & 0.020 & 0.904 & 0.051 & \textcolor{colMIS}{0.018} &  &  &  \\
 &  & 8 & 0.924 & 0.031 & \textcolor{colMIS}{0.019} & 0.970 & \textcolor{colCPD}{0.030} & 0.022 & 0.926 & \textcolor{colCPD}{0.030} & 0.022 &  &  &  \\
 &  & 9 & 0.943 & 0.018 & \textcolor{colMIS}{0.022} & 0.969 & 0.029 & 0.024 & 0.952 & \textcolor{colCPD}{0.015} & 0.029 &  &  &  \\
 &  & 10 & 0.931 & 0.028 & 0.083 & 0.968 & 0.028 & \textcolor{colMIS}{0.025} & 0.948 & \textcolor{colCPD}{0.014} & 0.132 &  &  &  \\
\cmidrule{3-15}
 & & Mean (1--10) & 0.919 & 0.036 & 0.021 & 0.972 & \textcolor{colCPD}{0.030} & \textcolor{colMIS}{0.019} & 0.923 & 0.034 & 0.027 &  &  &  \\
\cmidrule{2-15}
 & M & 1 & 0.941 & 0.017 & \textcolor{colMIS}{0.014} & 0.980 & 0.032 & 0.021 & 0.942 & 0.016 & \textcolor{colMIS}{0.014} & 0.956 & \textcolor{colCPD}{0.011} & 0.017 \\
 &  & 2 & 0.935 & 0.021 & \textcolor{colMIS}{0.014} & 0.981 & 0.033 & 0.021 & 0.936 & 0.020 & \textcolor{colMIS}{0.014} & 0.957 & \textcolor{colCPD}{0.011} & 0.018 \\
 &  & 3 & 0.931 & 0.025 & \textcolor{colMIS}{0.015} & 0.984 & 0.034 & 0.022 & 0.931 & 0.024 & \textcolor{colMIS}{0.015} & 0.957 & \textcolor{colCPD}{0.012} & 0.018 \\
 &  & 4 & 0.926 & 0.029 & \textcolor{colMIS}{0.016} & 0.987 & 0.037 & 0.023 & 0.927 & 0.029 & \textcolor{colMIS}{0.016} & 0.956 & \textcolor{colCPD}{0.014} & 0.019 \\
 &  & 5 & 0.921 & 0.033 & \textcolor{colMIS}{0.017} & 0.989 & 0.039 & 0.025 & 0.924 & 0.032 & \textcolor{colMIS}{0.017} & 0.952 & \textcolor{colCPD}{0.014} & 0.019 \\
\cmidrule{3-15}
 & & Mean (1--5) & 0.931 & 0.025 & \textcolor{colMIS}{0.015} & 0.984 & 0.035 & 0.023 & 0.932 & 0.024 & \textcolor{colMIS}{0.015} & 0.955 & \textcolor{colCPD}{0.012} & 0.018 \\
\cmidrule{3-15}
 &  & 6 & 0.916 & 0.039 & \textcolor{colMIS}{0.019} & 0.989 & 0.039 & 0.026 & 0.918 & \textcolor{colCPD}{0.037} & \textcolor{colMIS}{0.019} &  &  &  \\
 &  & 7 & 0.913 & 0.043 & 0.023 & 0.990 & \textcolor{colCPD}{0.040} & 0.027 & 0.914 & 0.041 & \textcolor{colMIS}{0.022} &  &  &  \\
 &  & 8 & 0.917 & 0.040 & 0.027 & 0.991 & 0.041 & 0.028 & 0.919 & \textcolor{colCPD}{0.038} & \textcolor{colMIS}{0.025} &  &  &  \\
 &  & 9 & 0.925 & 0.033 & 0.036 & 0.993 & 0.043 & \textcolor{colMIS}{0.029} & 0.931 & \textcolor{colCPD}{0.031} & 0.043 &  &  &  \\
 &  & 10 & 0.923 & 0.032 & 0.102 & 0.993 & 0.043 & \textcolor{colMIS}{0.030} & 0.947 & \textcolor{colCPD}{0.015} & 0.144 &  &  &  \\
\cmidrule{3-15}
 & & Mean (1--10) & 0.925 & 0.031 & 0.028 & 0.988 & 0.038 & \textcolor{colMIS}{0.025} & 0.929 & \textcolor{colCPD}{0.028} & 0.033 &  &  &  \\
\midrule
ETS & F & 1 & 0.935 & 0.022 & \textcolor{colMIS}{0.008} & 0.967 & 0.029 & 0.013 & 0.935 & 0.022 & \textcolor{colMIS}{0.008} & 0.956 & \textcolor{colCPD}{0.011} & 0.011 \\
 &  & 2 & 0.931 & 0.027 & \textcolor{colMIS}{0.009} & 0.969 & 0.029 & 0.014 & 0.932 & 0.026 & \textcolor{colMIS}{0.009} & 0.959 & \textcolor{colCPD}{0.015} & 0.012 \\
 &  & 3 & 0.926 & 0.030 & 0.010 & 0.972 & 0.029 & 0.015 & 0.927 & 0.029 & \textcolor{colMIS}{0.009} & 0.959 & \textcolor{colCPD}{0.016} & 0.012 \\
 &  & 4 & 0.924 & 0.032 & \textcolor{colMIS}{0.010} & 0.976 & 0.031 & 0.016 & 0.925 & 0.031 & \textcolor{colMIS}{0.010} & 0.965 & \textcolor{colCPD}{0.020} & 0.013 \\
 &  & 5 & 0.922 & 0.033 & \textcolor{colMIS}{0.010} & 0.977 & 0.032 & 0.018 & 0.924 & 0.032 & \textcolor{colMIS}{0.010} & 0.970 & \textcolor{colCPD}{0.024} & 0.014 \\
\cmidrule{3-15}
 & & Mean (1--5) & 0.928 & 0.029 & \textcolor{colMIS}{0.009} & 0.972 & 0.030 & 0.015 & 0.929 & 0.028 & \textcolor{colMIS}{0.009} & 0.962 & \textcolor{colCPD}{0.017} & 0.012 \\
\cmidrule{3-15}
 &  & 6 & 0.922 & 0.035 & 0.012 & 0.975 & \textcolor{colCPD}{0.031} & 0.020 & 0.924 & 0.033 & \textcolor{colMIS}{0.011} &  &  &  \\
 &  & 7 & 0.919 & 0.039 & \textcolor{colMIS}{0.014} & 0.971 & \textcolor{colCPD}{0.030} & 0.020 & 0.923 & 0.037 & 0.015 &  &  &  \\
 &  & 8 & 0.925 & 0.033 & \textcolor{colMIS}{0.015} & 0.970 & 0.031 & 0.021 & 0.931 & \textcolor{colCPD}{0.029} & 0.021 &  &  &  \\
 &  & 9 & 0.930 & 0.030 & \textcolor{colMIS}{0.021} & 0.970 & 0.030 & 0.022 & 0.939 & \textcolor{colCPD}{0.021} & 0.041 &  &  &  \\
 &  & 10 & 0.906 & 0.049 & 0.060 & 0.967 & 0.025 & \textcolor{colMIS}{0.023} & 0.951 & \textcolor{colCPD}{0.016} & 0.161 &  &  &  \\
\cmidrule{3-15}
 & & Mean (1--10) & 0.924 & 0.033 & \textcolor{colMIS}{0.017} & 0.971 & 0.030 & 0.018 & 0.931 & \textcolor{colCPD}{0.028} & 0.029 &  &  &  \\
\cmidrule{2-15}
 & M & 1 & 0.945 & 0.015 & \textcolor{colMIS}{0.014} & 0.979 & 0.031 & 0.020 & 0.946 & 0.014 & \textcolor{colMIS}{0.014} & 0.956 & \textcolor{colCPD}{0.010} & 0.017 \\
 &  & 2 & 0.943 & 0.017 & \textcolor{colMIS}{0.014} & 0.980 & 0.032 & 0.021 & 0.944 & 0.017 & \textcolor{colMIS}{0.014} & 0.956 & \textcolor{colCPD}{0.011} & 0.017 \\
 &  & 3 & 0.940 & 0.019 & \textcolor{colMIS}{0.015} & 0.982 & 0.033 & 0.022 & 0.940 & 0.019 & \textcolor{colMIS}{0.015} & 0.958 & \textcolor{colCPD}{0.011} & 0.018 \\
 &  & 4 & 0.940 & 0.022 & \textcolor{colMIS}{0.016} & 0.984 & 0.035 & 0.023 & 0.941 & 0.021 & \textcolor{colMIS}{0.016} & 0.957 & \textcolor{colCPD}{0.012} & 0.019 \\
 &  & 5 & 0.939 & 0.024 & \textcolor{colMIS}{0.017} & 0.985 & 0.036 & 0.024 & 0.941 & 0.023 & \textcolor{colMIS}{0.017} & 0.956 & \textcolor{colCPD}{0.015} & 0.020 \\
\cmidrule{3-15}
 & & Mean (1--5) & 0.941 & 0.019 & \textcolor{colMIS}{0.015} & 0.982 & 0.033 & 0.022 & 0.942 & 0.019 & \textcolor{colMIS}{0.015} & 0.957 & \textcolor{colCPD}{0.012} & 0.018 \\
\cmidrule{3-15}
 &  & 6 & 0.948 & 0.020 & \textcolor{colMIS}{0.017} & 0.987 & 0.037 & 0.026 & 0.950 & \textcolor{colCPD}{0.019} & 0.018 &  &  &  \\
 &  & 7 & 0.951 & \textcolor{colCPD}{0.022} & \textcolor{colMIS}{0.019} & 0.987 & 0.037 & 0.026 & 0.951 & \textcolor{colCPD}{0.022} & \textcolor{colMIS}{0.019} &  &  &  \\
 &  & 8 & 0.956 & \textcolor{colCPD}{0.021} & \textcolor{colMIS}{0.022} & 0.987 & 0.038 & 0.027 & 0.958 & \textcolor{colCPD}{0.021} & 0.023 &  &  &  \\
 &  & 9 & 0.962 & \textcolor{colCPD}{0.020} & 0.037 & 0.990 & 0.040 & \textcolor{colMIS}{0.027} & 0.965 & 0.021 & 0.047 &  &  &  \\
 &  & 10 & 0.931 & 0.033 & 0.114 & 0.994 & 0.044 & \textcolor{colMIS}{0.028} & 0.960 & \textcolor{colCPD}{0.015} & 0.235 &  &  &  \\
\cmidrule{3-15}
 & & Mean (1--10) & 0.945 & 0.021 & 0.029 & 0.986 & 0.036 & \textcolor{colMIS}{0.025} & 0.950 & \textcolor{colCPD}{0.019} & 0.042 &  &  &  \\
\bottomrule
\end{longtable}
\end{small}
\end{center}

\vspace{-.4in}

Adaptive conformal prediction performs the best for $h=1,\dots,5$, the horizons for which it is computed. It attains the smallest CPD at each of the horizons for both sexes and forecasting methods, and its ECP is at or slightly above the nominal level throughout, ranging from 0.952 to 0.970. Its advantage holds as the horizon increases. While the under-coverage of the split and non-conformity score predictions worsens with $h$, the CPD of adaptive conformal prediction remains small, between 0.010 and 0.024, due to the continual adaptation of the significance level.

The MIS balances the width of an interval against a penalty for the observations that fall outside it. Since a narrow interval that under-covers can still attain a small MIS, we regard the coverage as the primary criterion and use the MIS to further compare methods whose CPDs are similar. At short horizons, the four methods have similar MIS. At the longest horizons, however, the sequential predictive conformal prediction attains the smallest MIS. This is mainly because, at longer horizons, the split and non-conformity score intervals are calibrated on only a few validation forecasts, which makes them unstable and wider.

\begin{figure}[!htbp]
\centering
\includegraphics[width=1\textwidth]{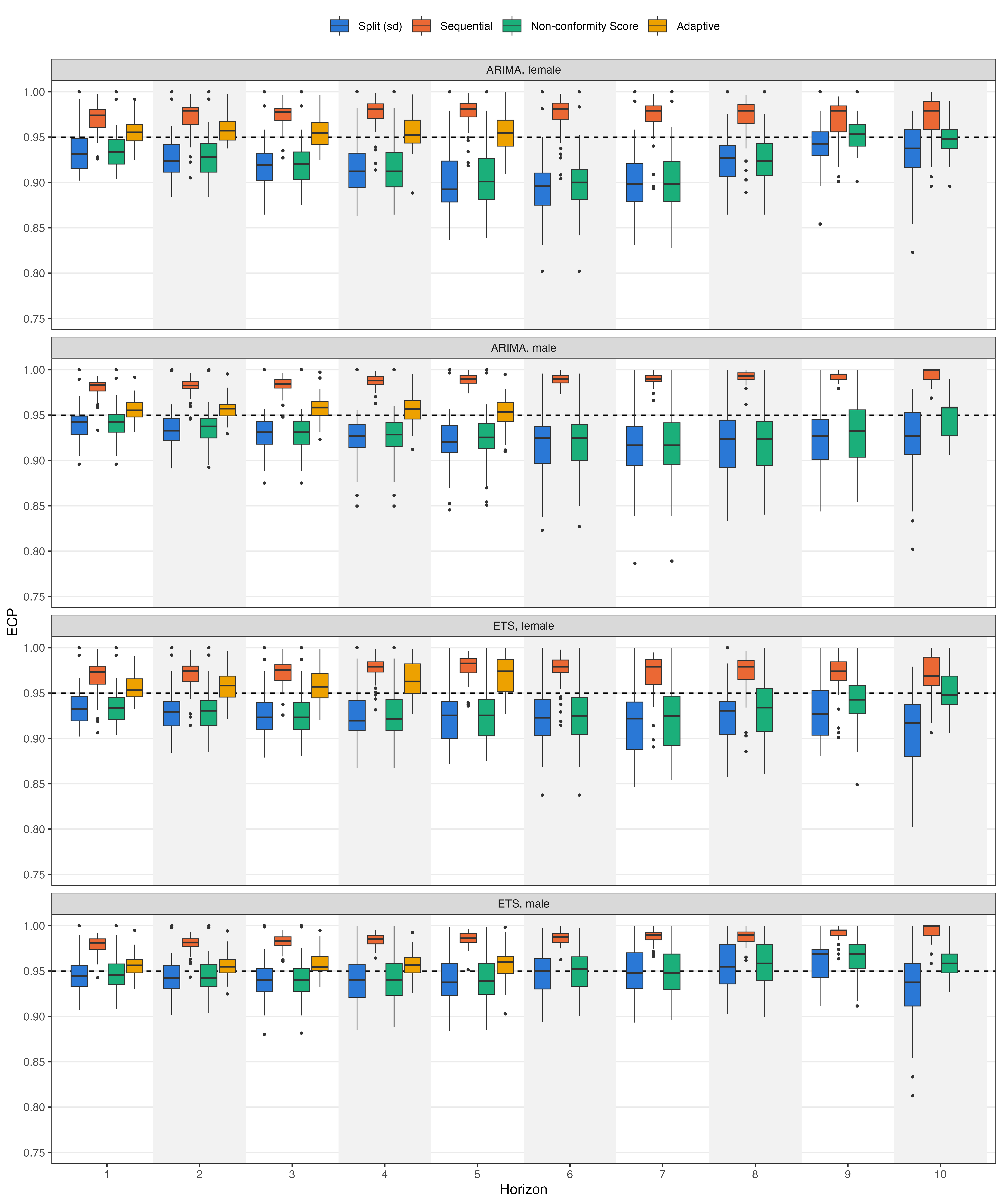}
\caption{\small One- to ten-step-ahead ECPs of the 47 prefectures in Japan for the four conformal prediction methods at the nominal coverage probability of 95\% (dashed line), for each combination of forecasting method and sex. Adaptive conformal prediction is computed for $h=1,\dots,5$ only.}\label{fig:6}
\end{figure}

We display one- to ten-step-ahead ECPs for the four conformal prediction methods in Figure~\ref{fig:6}. For each combination of forecasting method and sex, we present side-by-side boxplots of the ECPs of the 47 prefectures at each horizon. For the split and non-conformity score conformal prediction, there is a tendency to under-cover the coverage for female, especially from $h=3$ to $h=7$. Such under-coverage likely arises because the calibration on the validation set does not fully carry over to the test set. In contrast, the sequential predictive conformal prediction often over-covers the coverage and tends to be conservative. Among all conformal prediction methods, the adaptive conformal prediction, computed for $h=1,\dots,5$ only, yields ECPs centered closest to the nominal level, with a smaller spread across prefectures than split and non-conformity score predictions.

\begin{figure}[!htbp]
\centering
\includegraphics[width=1\textwidth]{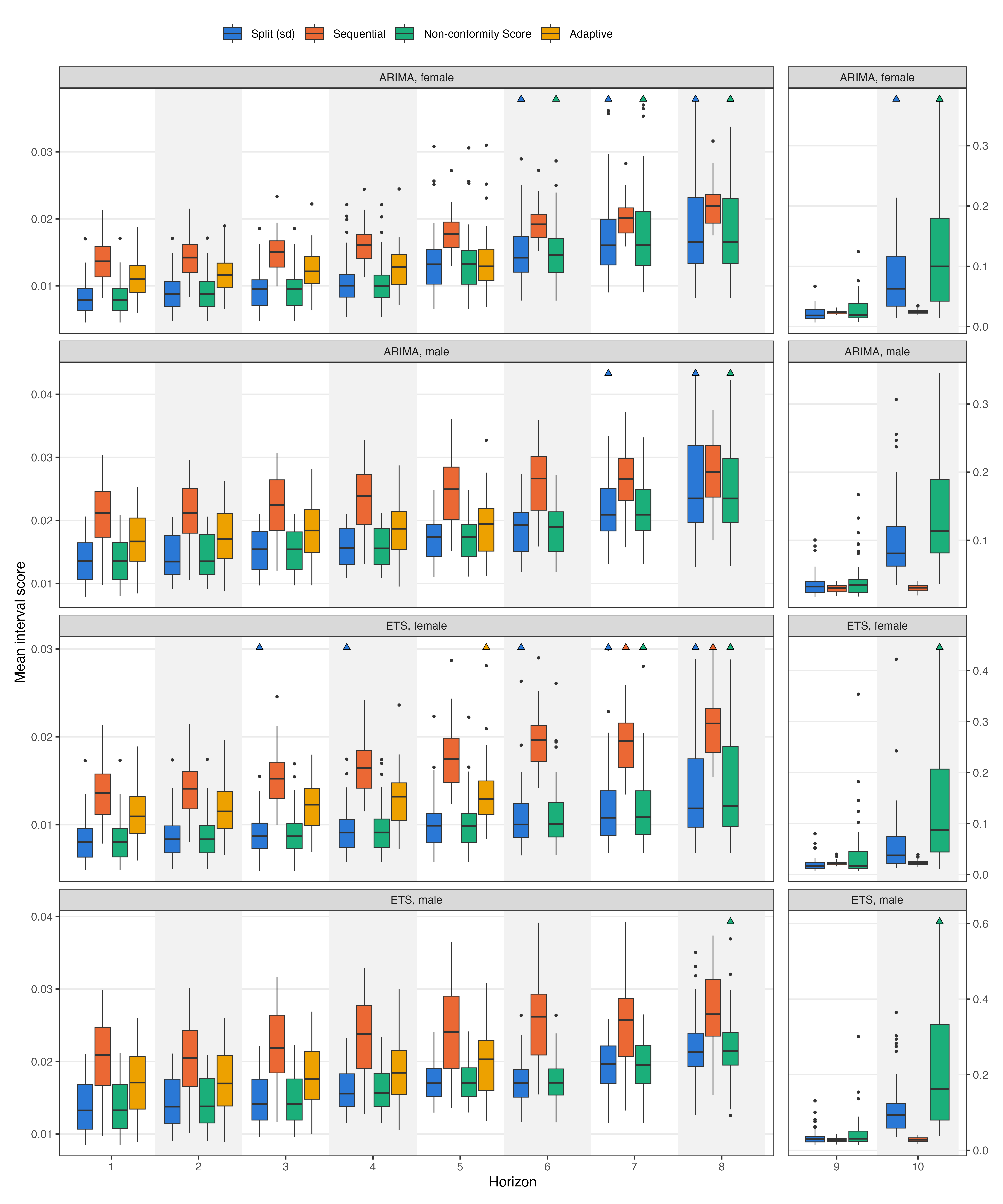}
\caption{\small One- to ten-step-ahead MISs of the 47 prefectures in Japan for the four conformal prediction methods at the nominal coverage probability of 95\%, for each combination of forecasting method and sex. Horizons $h=1,\dots,8$ and $h=9,10$ are shown with separate vertical axes, on the left and right, respectively, as the MISs at $h=9$ and 10 are on a much larger scale. Each vertical axis ends at the top of the highest box or whisker; values above it are shown as triangles. Adaptive conformal prediction is computed for $h=1,\dots,5$ only.}\label{fig:7}
\end{figure}

In Figure~\ref{fig:7}, we further present one- to ten-step-ahead MIS for all four methods. At short horizons, the MISs of the four methods are comparable, as all are calibrated on enough validation errors, and split conformal and non-conformity score predictions attain the smallest MISs. The differences are largest at longer horizons, especially at $h=10$, where the MISs of split conformal and non-conformity score predictions increase sharply and vary widely across prefectures, whereas those of sequential predictive conformal prediction remain small. This is because split conformal and non-conformity score predictions calibrate their intervals on far fewer forecast errors in the validation set when $h$ increases, which makes the calibrated multipliers, and hence the interval widths, unstable. In contrast, sequential predictive conformal prediction re-estimates its interval at each forecast origin from all earlier forecast errors. At longer horizons, the large MISs of these two methods therefore arise mainly from their wide intervals. The MIS is therefore most useful for distinguishing conformal prediction methods whose ECPs and CPDs are similar.

\section{Conclusion}\label{sec:7}

We present conformal prediction as a way to quantify forecast uncertainty for high-dimensional functional time series and illustrate it with Japanese subnational age- and sex-specific log mortality rates from 1975 to 2023. The conformal prediction framework is a model-agnostic, distribution-free means of constructing prediction intervals. Using a one-way functional analysis of variance and a functional factor model, we evaluate and compare the finite-sample performance of split, non-conformity score, sequential predictive, and adaptive conformal predictions. To facilitate reproducibility, the \Rlogo \ code for computing the interval forecast errors obtained from the split, non-conformity score, sequential predictive, and adaptive conformal predictions using the Japanese and Canadian mortality data is available at \url{https://github.com/hanshang/conformal_prediction_OWFANOVA_FFM}.

Split conformal, non-conformity score, and adaptive conformal predictions require sample splitting, whereas sequential predictive conformal prediction does not. Split conformal and non-conformity score predictions perform similarly at short horizons, but the latter is computationally more efficient as it requires no tuning. Adaptive conformal prediction requires an additional set to tune its learning rate, which restricts it to shorter horizons when the data are not long enough. In terms of accuracy, adaptive conformal prediction attains the smallest CPD at horizons one to five in most settings. At longer horizons, sequential predictive conformal prediction is recommended: at the nominal coverage probability of 95\%, its CPD is comparable to those of split conformal and non-conformity score predictions, but its MIS is much smaller, indicating the sharper intervals with similar coverage.

There are several ways in which the current paper may be further extended, and we briefly outline four:
\begin{inparaenum}
\item[1)] In the sequential predictive conformal prediction, we model the temporal dependence of the absolute residuals via an autoregressive process in a quantile regression. Other time-series models can also be applied.
\item[2)] Due to sex differences, the female and male data are modeled separately. For each sex, the observations form a three-dimensional array indexed by age, calendar year, and prefecture. Modeling both sexes jointly increases the effective dimensionality of the problem, thereby requiring more sophisticated statistical tools to decompose and analyze such data. In this setting, approaches such as the two-way functional analysis of variance of \cite{JSS24}, coupled with functional factor models, provide natural and flexible frameworks.
\item[3)] The conformal prediction developed can also be extended to quantify forecast uncertainty for subnational fertility rates \citep[see, e.g.,][]{YSR24}.
\item[4)] In our data splitting scheme, the number of validation forecast errors decreases with the horizon. A different splitting scheme that gives the same calibration length for all horizons could be tested, which may stabilize the intervals at longer horizons.
\end{inparaenum}

\section*{Acknowledgements}

The first author thanks the financial support from an Australian Research Council Discovery Project (DP230102250), an Australian Research Council Future Fellowship (FT240100338), and a National Natural Science Foundation of China General Program (NSFC72673148). The second author gratefully acknowledges the financial support provided through the International Macquarie University Research Excellence Scholarship. 

\section*{Disclosure statement}

No potential conflict of interest was reported by the authors.

\appendix
\section*{Appendix: Canadian subnational age- and sex-specific mortality rates}

As a sensitivity analysis, we also consider age-specific $\log$ mortality data for 12 Canadian provinces and territories, obtained from the Canadian Human Mortality Database (\url{http://www.bdlc.umontreal.ca/chmd/}). This database has no data for the Nunavut territory. For most regions, observations are available through 2023, except for Yukon, where the data end in 2016. For consistency and completeness across regions, we restrict the study period to 1950--2016. To mitigate the impact of measurement error at advanced ages, ages are grouped from 0 to 79 in single-year intervals, with ages 80 and above aggregated into an open-ended age group. These data do not constitute an HDFTS, since $N=12$ and $T = 67$, so that $N<T$; nevertheless, the proposed method can still be applied.

The initial training data are from 1950 to 1995; the validation data are from 1996 to 2006; and the test data are from 2007 to 2016. For the adaptive conformal prediction, the calibration set consists of 11 years from 1985 to 1995, the last 11 years of the initial training data, and the validation data are used to choose the learning rate $\lambda_s$; as for the Japanese data, it is applied for $h=1,\dots,5$ only. Using the same forecast scheme as in Section~\ref{sec:5}, with the error measure adapted to the different number of ages, we apply split, non-conformity score, sequential predictive, and adaptive conformal predictions to evaluate and compare the accuracy of interval forecasts in Table~\ref{tab:2}.
\begin{center}
\tabcolsep 0.05in
\renewcommand{\arraystretch}{0.96}
\begin{small}
\begin{longtable}{@{}lllrrrrrrrrrrrr@{}}
\caption{\small Averaged over 12 provinces and territories and ages in Canada, the one- to ten-step-ahead interval forecast accuracy is presented for the conformal prediction methods, namely split, sequential, non-conformity score, and adaptive conformal predictions, at the nominal coverage probability of 95\%. In split conformal prediction, the interval is based on the standard deviation (sd) of the validation residuals. The adaptive conformal prediction is computed for $h=1,\dots,5$ only; the mean over $h=1,\dots,5$ compares all four methods, and the mean over $h=1,\dots,10$ the other three. For forecasting the estimated factors, we consider two univariate time-series forecasting methods, namely ARIMA and ETS. In each row, the smallest CPD and the smallest MIS are shown in red and blue, respectively.}\label{tab:2} \\
\toprule
 & & & \multicolumn{3}{c}{Split (sd)} & \multicolumn{3}{c}{Sequential} & \multicolumn{3}{c}{\begin{tabular}[b]{@{}c@{}}Non-conformity\\ Score\end{tabular}} & \multicolumn{3}{c}{Adaptive} \\
\cmidrule(lr){4-6}\cmidrule(lr){7-9}\cmidrule(lr){10-12}\cmidrule(lr){13-15}
Method & Sex & $h$ & ECP & CPD & MIS & ECP & CPD & MIS & ECP & CPD & MIS & ECP & CPD & MIS \\
\midrule
\endfirsthead
\toprule
 & & & \multicolumn{3}{c}{Split (sd)} & \multicolumn{3}{c}{Sequential} & \multicolumn{3}{c}{\begin{tabular}[b]{@{}c@{}}Non-conformity\\ Score\end{tabular}} & \multicolumn{3}{c}{Adaptive} \\
\cmidrule(lr){4-6}\cmidrule(lr){7-9}\cmidrule(lr){10-12}\cmidrule(lr){13-15}
Method & Sex & $h$ & ECP & CPD & MIS & ECP & CPD & MIS & ECP & CPD & MIS & ECP & CPD & MIS \\
\midrule
\endhead
\midrule
\multicolumn{15}{r}{Continued on next page} \\
\endfoot
\endlastfoot
ARIMA & F & 1 & 0.938 & 0.012 & 0.033 & 0.992 & 0.043 & \textcolor{colMIS}{0.023} & 0.939 & 0.012 & 0.039 & 0.950 & \textcolor{colCPD}{0.007} & 0.043 \\
 &  & 2 & 0.930 & 0.021 & 0.031 & 0.993 & 0.043 & \textcolor{colMIS}{0.023} & 0.933 & 0.018 & 0.037 & 0.949 & \textcolor{colCPD}{0.006} & 0.040 \\
 &  & 3 & 0.926 & 0.024 & 0.031 & 0.993 & 0.043 & \textcolor{colMIS}{0.024} & 0.929 & 0.022 & 0.037 & 0.944 & \textcolor{colCPD}{0.008} & 0.040 \\
 &  & 4 & 0.922 & 0.028 & 0.031 & 0.994 & 0.044 & \textcolor{colMIS}{0.024} & 0.927 & 0.024 & 0.046 & 0.946 & \textcolor{colCPD}{0.011} & 0.039 \\
 &  & 5 & 0.918 & 0.032 & 0.033 & 0.994 & 0.044 & \textcolor{colMIS}{0.024} & 0.926 & 0.025 & 0.052 & 0.942 & \textcolor{colCPD}{0.019} & 0.049 \\
\cmidrule{3-15}
 & & Mean (1--5) & 0.927 & 0.024 & 0.032 & 0.993 & 0.043 & \textcolor{colMIS}{0.023} & 0.931 & 0.020 & 0.042 & 0.946 & \textcolor{colCPD}{0.010} & 0.042 \\
\cmidrule{3-15}
 &  & 6 & 0.911 & 0.039 & 0.036 & 0.994 & 0.044 & \textcolor{colMIS}{0.024} & 0.921 & \textcolor{colCPD}{0.030} & 0.057 &  &  &  \\
 &  & 7 & 0.914 & 0.037 & 0.036 & 0.996 & 0.046 & \textcolor{colMIS}{0.024} & 0.925 & \textcolor{colCPD}{0.027} & 0.055 &  &  &  \\
 &  & 8 & 0.918 & 0.034 & 0.063 & 0.994 & 0.044 & \textcolor{colMIS}{0.024} & 0.935 & \textcolor{colCPD}{0.020} & 0.108 &  &  &  \\
 &  & 9 & 0.884 & 0.066 & 0.035 & 0.995 & 0.045 & \textcolor{colMIS}{0.025} & 0.942 & \textcolor{colCPD}{0.014} & 0.177 &  &  &  \\
 &  & 10 & 0.847 & 0.108 & 0.074 & 0.994 & 0.044 & \textcolor{colMIS}{0.025} & 0.939 & \textcolor{colCPD}{0.019} & 0.791 &  &  &  \\
\cmidrule{3-15}
 & & Mean (1--10) & 0.911 & 0.040 & 0.040 & 0.994 & 0.044 & \textcolor{colMIS}{0.024} & 0.931 & \textcolor{colCPD}{0.021} & 0.140 &  &  &  \\
\cmidrule{2-15}
 & M & 1 & 0.941 & 0.015 & 0.036 & 0.990 & 0.042 & \textcolor{colMIS}{0.019} & 0.944 & 0.013 & 0.044 & 0.950 & \textcolor{colCPD}{0.007} & 0.029 \\
 &  & 2 & 0.937 & 0.017 & 0.037 & 0.991 & 0.043 & \textcolor{colMIS}{0.020} & 0.939 & 0.016 & 0.042 & 0.951 & \textcolor{colCPD}{0.009} & 0.029 \\
 &  & 3 & 0.937 & 0.018 & 0.035 & 0.992 & 0.042 & \textcolor{colMIS}{0.020} & 0.940 & 0.017 & 0.040 & 0.954 & \textcolor{colCPD}{0.009} & 0.031 \\
 &  & 4 & 0.940 & 0.018 & 0.038 & 0.991 & 0.041 & \textcolor{colMIS}{0.020} & 0.941 & 0.016 & 0.046 & 0.951 & \textcolor{colCPD}{0.015} & 0.034 \\
 &  & 5 & 0.943 & 0.021 & 0.036 & 0.991 & 0.041 & \textcolor{colMIS}{0.021} & 0.949 & 0.015 & 0.047 & 0.954 & \textcolor{colCPD}{0.010} & 0.032 \\
\cmidrule{3-15}
 & & Mean (1--5) & 0.940 & 0.018 & 0.036 & 0.991 & 0.042 & \textcolor{colMIS}{0.020} & 0.943 & 0.015 & 0.044 & 0.952 & \textcolor{colCPD}{0.010} & 0.031 \\
\cmidrule{3-15}
 &  & 6 & 0.938 & 0.022 & 0.037 & 0.991 & 0.041 & \textcolor{colMIS}{0.021} & 0.945 & \textcolor{colCPD}{0.015} & 0.049 &  &  &  \\
 &  & 7 & 0.939 & 0.029 & 0.487 & 0.992 & 0.042 & \textcolor{colMIS}{0.021} & 0.945 & \textcolor{colCPD}{0.016} & 0.051 &  &  &  \\
 &  & 8 & 0.944 & 0.023 & 0.053 & 0.993 & 0.043 & \textcolor{colMIS}{0.021} & 0.953 & \textcolor{colCPD}{0.018} & 0.077 &  &  &  \\
 &  & 9 & 0.920 & 0.047 & 0.052 & 0.991 & 0.041 & \textcolor{colMIS}{0.021} & 0.947 & \textcolor{colCPD}{0.023} & 0.140 &  &  &  \\
 &  & 10 & 0.872 & 0.084 & 0.093 & 0.988 & 0.040 & \textcolor{colMIS}{0.022} & 0.956 & \textcolor{colCPD}{0.014} & 0.310 &  &  &  \\
\cmidrule{3-15}
 & & Mean (1--10) & 0.931 & 0.029 & 0.090 & 0.991 & 0.042 & \textcolor{colMIS}{0.021} & 0.946 & \textcolor{colCPD}{0.016} & 0.085 &  &  &  \\
\midrule
ETS & F & 1 & 0.937 & 0.014 & 0.031 & 0.991 & 0.042 & \textcolor{colMIS}{0.022} & 0.940 & 0.013 & 0.038 & 0.950 & \textcolor{colCPD}{0.008} & 0.043 \\
 &  & 2 & 0.932 & 0.018 & 0.031 & 0.992 & 0.042 & \textcolor{colMIS}{0.022} & 0.937 & 0.015 & 0.036 & 0.949 & \textcolor{colCPD}{0.007} & 0.038 \\
 &  & 3 & 0.930 & 0.020 & 0.030 & 0.991 & 0.043 & \textcolor{colMIS}{0.023} & 0.934 & 0.018 & 0.036 & 0.942 & \textcolor{colCPD}{0.010} & 0.036 \\
 &  & 4 & 0.930 & 0.020 & 0.031 & 0.991 & 0.043 & \textcolor{colMIS}{0.024} & 0.936 & 0.016 & 0.046 & 0.946 & \textcolor{colCPD}{0.011} & 0.036 \\
 &  & 5 & 0.927 & 0.023 & 0.034 & 0.992 & 0.044 & \textcolor{colMIS}{0.025} & 0.935 & \textcolor{colCPD}{0.017} & 0.053 & 0.941 & 0.020 & 0.037 \\
\cmidrule{3-15}
 & & Mean (1--5) & 0.932 & 0.019 & 0.031 & 0.991 & 0.043 & \textcolor{colMIS}{0.023} & 0.937 & 0.015 & 0.042 & 0.946 & \textcolor{colCPD}{0.011} & 0.038 \\
\cmidrule{3-15}
 &  & 6 & 0.927 & 0.024 & 0.037 & 0.993 & 0.044 & \textcolor{colMIS}{0.025} & 0.934 & \textcolor{colCPD}{0.018} & 0.057 &  &  &  \\
 &  & 7 & 0.926 & 0.028 & 0.036 & 0.994 & 0.045 & \textcolor{colMIS}{0.025} & 0.936 & \textcolor{colCPD}{0.019} & 0.055 &  &  &  \\
 &  & 8 & 0.927 & 0.027 & 0.072 & 0.997 & 0.047 & \textcolor{colMIS}{0.025} & 0.943 & \textcolor{colCPD}{0.014} & 0.125 &  &  &  \\
 &  & 9 & 0.889 & 0.061 & 0.036 & 0.993 & 0.043 & \textcolor{colMIS}{0.026} & 0.950 & \textcolor{colCPD}{0.011} & 0.157 &  &  &  \\
 &  & 10 & 0.889 & 0.063 & 0.167 & 0.995 & 0.045 & \textcolor{colMIS}{0.026} & 0.947 & \textcolor{colCPD}{0.015} & 1.491 &  &  &  \\
\cmidrule{3-15}
 & & Mean (1--10) & 0.922 & 0.030 & 0.050 & 0.993 & 0.044 & \textcolor{colMIS}{0.024} & 0.939 & \textcolor{colCPD}{0.015} & 0.209 &  &  &  \\
\cmidrule{2-15}
 & M & 1 & 0.939 & 0.019 & 0.030 & 0.990 & 0.043 & \textcolor{colMIS}{0.019} & 0.945 & 0.013 & 0.042 & 0.951 & \textcolor{colCPD}{0.007} & 0.029 \\
 &  & 2 & 0.939 & 0.015 & 0.038 & 0.991 & 0.043 & \textcolor{colMIS}{0.020} & 0.940 & 0.015 & 0.040 & 0.951 & \textcolor{colCPD}{0.009} & 0.030 \\
 &  & 3 & 0.939 & 0.019 & 0.035 & 0.993 & 0.043 & \textcolor{colMIS}{0.020} & 0.942 & 0.017 & 0.040 & 0.953 & \textcolor{colCPD}{0.010} & 0.031 \\
 &  & 4 & 0.942 & 0.018 & 0.041 & 0.992 & 0.043 & \textcolor{colMIS}{0.020} & 0.945 & 0.016 & 0.048 & 0.950 & \textcolor{colCPD}{0.015} & 0.030 \\
 &  & 5 & 0.942 & 0.023 & 0.036 & 0.991 & 0.041 & \textcolor{colMIS}{0.020} & 0.949 & 0.017 & 0.048 & 0.954 & \textcolor{colCPD}{0.012} & 0.031 \\
\cmidrule{3-15}
 & & Mean (1--5) & 0.940 & 0.019 & 0.036 & 0.992 & 0.043 & \textcolor{colMIS}{0.020} & 0.944 & 0.016 & 0.044 & 0.952 & \textcolor{colCPD}{0.011} & 0.030 \\
\cmidrule{3-15}
 &  & 6 & 0.939 & 0.024 & 0.037 & 0.993 & 0.043 & \textcolor{colMIS}{0.020} & 0.947 & \textcolor{colCPD}{0.016} & 0.050 &  &  &  \\
 &  & 7 & 0.930 & 0.029 & 0.034 & 0.993 & 0.043 & \textcolor{colMIS}{0.021} & 0.942 & \textcolor{colCPD}{0.018} & 0.051 &  &  &  \\
 &  & 8 & 0.947 & 0.025 & 0.053 & 0.993 & 0.043 & \textcolor{colMIS}{0.021} & 0.952 & \textcolor{colCPD}{0.016} & 0.073 &  &  &  \\
 &  & 9 & 0.918 & 0.050 & 0.052 & 0.992 & 0.042 & \textcolor{colMIS}{0.021} & 0.949 & \textcolor{colCPD}{0.024} & 0.163 &  &  &  \\
 &  & 10 & 0.883 & 0.074 & 0.094 & 0.988 & 0.038 & \textcolor{colMIS}{0.021} & 0.959 & \textcolor{colCPD}{0.011} & 0.505 &  &  &  \\
\cmidrule{3-15}
 & & Mean (1--10) & 0.932 & 0.030 & 0.045 & 0.992 & 0.042 & \textcolor{colMIS}{0.020} & 0.947 & \textcolor{colCPD}{0.016} & 0.106 &  &  &  \\
\bottomrule
\end{longtable}
\end{small}
\end{center}

\vspace{-.3in}

\begin{figure}[!htbp]
\centering
\includegraphics[width=0.98\textwidth]{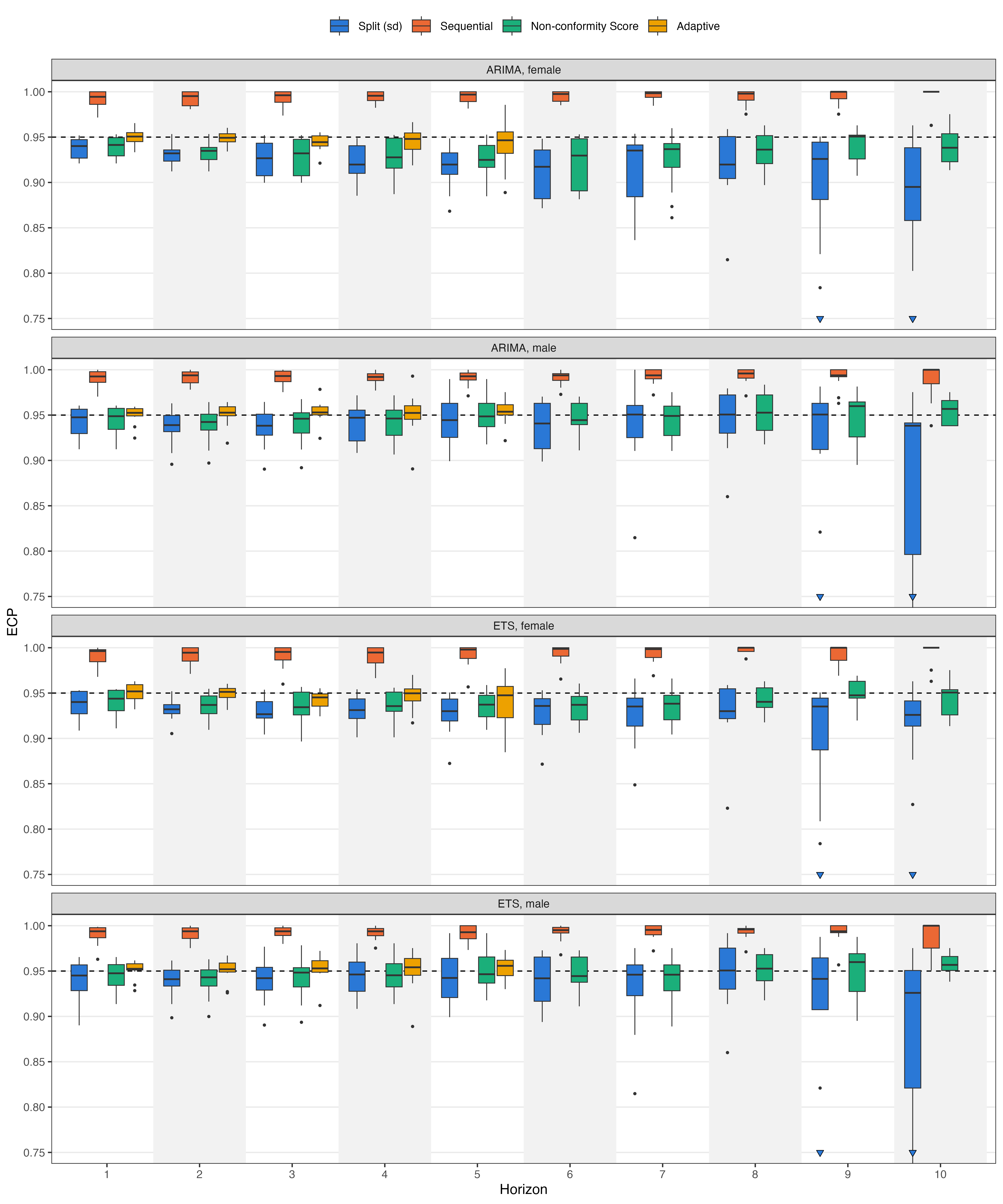}
\caption{\small One- to ten-step-ahead ECPs of 12 provinces and territories in Canada for the four conformal prediction methods at the nominal coverage probability of 95\% (dashed line), for each combination of forecasting method and sex. Adaptive conformal prediction is computed for $h=1,\dots,5$ only.}\label{fig:8}
\end{figure}

\begin{figure}[!htbp]
\centering
\includegraphics[width=0.98\textwidth]{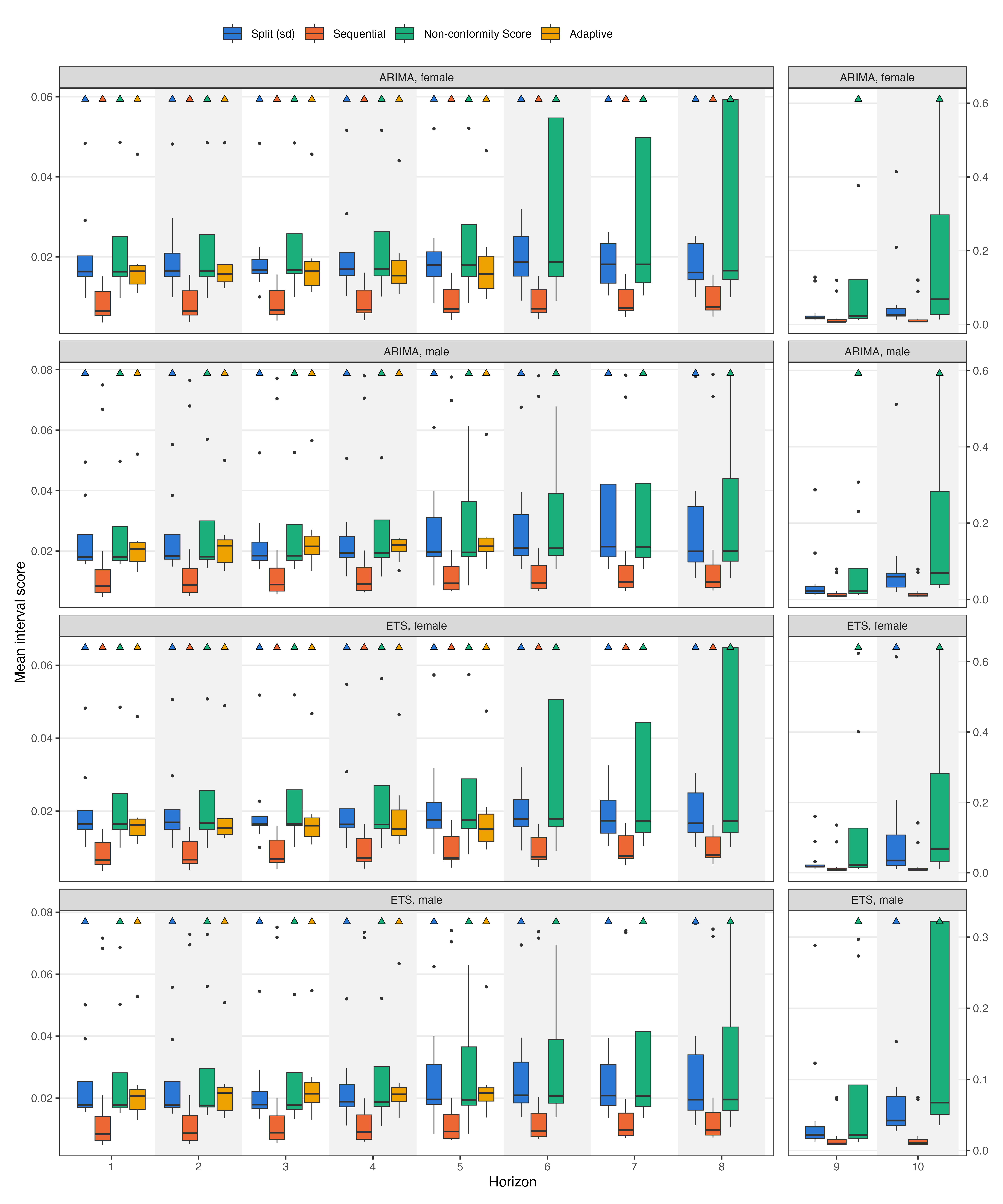}
\caption{\small One- to ten-step-ahead MISs of the 12 provinces and territories in Canada for the four conformal prediction methods at the nominal coverage probability of 95\%, for each combination of forecasting method and sex. Horizons $h=1,\dots,8$ and $h=9,10$ are shown with separate vertical axes, on the left and right, respectively, as the MISs at $h=9$ and 10 are on a much larger scale. Each vertical axis ends at the top of the highest box or whisker; values above it are shown as triangles. Adaptive conformal prediction is computed for $h=1,\dots,5$ only.}\label{fig:9}
\end{figure}

The detailed distributions of ECPs and MISs for different point forecast methods and sexes are presented as in Figures~\ref{fig:8} and~\ref{fig:9}. Consistent with earlier findings, split conformal and non-conformity score predictions tend to under-cover, whereas sequential predictive conformal prediction tends to over-cover. Adaptive conformal prediction attains the smallest CPD at almost all horizons from one to five, and is thus preferred at short horizons. At longer horizons, sequential predictive conformal prediction is recommended: its CPD is comparable to that of split conformal prediction, although larger than that of non-conformity score prediction, while its MIS is much smaller.

\newpage
\bibliographystyle{agsm}
\bibliography{conformal_HDFTS}

\end{document}